\documentclass[twocolumn]{aa} % for a referee version
\usepackage{graphicx}
\usepackage{natbib}
\usepackage{txfonts}
\usepackage{color}
\usepackage{hyperref}
\newcommand{\ecs}{erg cm$^{-3}$ s$^{-1}$}
\newcommand{\cmt}{cm$^{-3}$}
\newcommand{\cms}{cm$^{-2}$}
\newcommand{\modotyr}{M$_\odot$~yr$^{-1}$}

\usepackage{epsfig}

\begin{document}
   \title{Diagnostics of the molecular component of PDRs with Mechanical Heating}

   \subtitle{}

   \author{M. V. Kazandjian\inst{1}, R. Meijerink\inst{1, 2}, I. Pelupessy\inst{1}, F. P. Israel\inst{1}, M. Spaans\inst{2}}

   \institute{Sterrewacht Leiden, PO Box 9513, 2300 RA Leiden, The Netherlands\\
     \email{mher@strw.leidenuniv.nl}
       \and 
       Kapteyn Astronomical Institute, PO Box 800, 9700 AV Groningen, The Netherlands
     }

   \date{Received Xxxxxxxxx xx, xxxx; accepted Xxxxx xx, xxxx}

% \abstract{}{}{}{}{} 
% 5 {} token are mandatory
 
  \abstract
  % context heading (optional)
  % {} leave it empty if necessary 
{
  Multitransition CO observations of galaxy centers have revealed that
  significant fractions of the dense circumnuclear gas have high
  kinetic temperatures, which are hard to explain by pure photon
  excitation, but may be caused by dissipation of turbulent energy.
}
  % aims heading (mandatory) 
{
  We aim to determine to what extent mechanical heating should be taken into
  account while modelling PDRs. To this end, the effect of dissipated
  turbulence on the thermal and chemical properties of PDRs is explored.
}
  % methods heading (mandatory) 
{
Clouds are modelled as 1D
semi-infinite slabs whose thermal and chemical equilibrium is solved
for using the Leiden PDR-XDR code, where mechanical heating is added
as a constant term throughout the cloud.  An extensive parameter space
in hydrogen gas density, FUV radiation field and mechanical heating rate is
considered, covering almost all possible cases for the ISM relevant to
the conditions that are encountered in galaxies. Effects of mechanical
heating on the temperature profiles, column densities of CO and H$_2$O
and column density ratios of HNC, HCN and HCO$^+$ are discussed. 
}
  % results heading (mandatory)
{
  In a {{\it steady-state}} treatment, mechanical heating seems to play an
  important role in determining the kinetic temperature of the gas in
  molecular clouds. Particularly in high-energy
  environments such as starburst galaxies and galaxy centers, model
  gas temperatures are underestimated by at least a factor of
  two if mechanical heating is ignored. The models also show that CO, HCN
  and H$_2$O column densities increase as a function of mechanical heating. The
  HNC/HCN integrated column density ratio shows a decrease by a factor
  of at least two in high density regions with $n \sim 10^5$\,\cmt,
  whereas that of HCN/HCO$^+$ shows a strong dependence on mechanical
  heating for this same density range, with boosts of up to three
  orders of magnitude.
}
  % conclusions heading (optional), leave it empty if necessary 
{ 
The effects of mechanical heating cannot be ignored in studies of the
molecular gas excitation whenever the ratio of the star formation rate
to the gas density (SFR / $n^{3/2}$) is close to, or exceeds, $7
\times 10^{-6}$ M$_\odot$~yr$^{-1}$ cm$^{4.5}$~.  If mechanical
heating is not included, predicted column densities (such as those of
CO) are underestimated, sometimes (as in the case of HCN and HCO$^{+}$)
even by a few orders of magnitude.  As a lower bound to its
importance, we determined that it has non-negligible effects already
when mechanical heating is as little as 1\% of the UV heating in a PDR.
}


\authorrunning{Kazandjian {\it et. al}}
\titlerunning{Diagnostics of PDRs with Mechanical Heating}
\maketitle

\section{Introduction}

Radiation originating from the molecular gas in various galaxy
environments, including galaxy centers, provides information about the
physical state of these environments such as the gas mass and
temperature, or the source of its excitation. The far-ultraviolet
radiation (FUV; $6.0 < E < 13.6$~eV) emitted by newly formed luminous
stars and X-rays ($E > 1$~keV) produced by black hole accretion have
very distinct effects on the thermo-chemical balance of the
gas. Strong FUV illumination results in so-called photon-dominated
regions (PDRs; \cite{tielenshb1985}), while X-ray irradiation creates
X-ray dominated regions (XDRs; \cite{maloney96}). In PDRs and XDRs,
the thermal and chemical structure is completely determined by the
radiation field, and directly reflects the energy balance of the
interstellar gas. For example, in PDRs the CO is generally much cooler
(T$\sim$ 10-30~K) than in XDRs ($T\sim 20 - 500$~K), therefore XDRs 
generally exhibit CO emissions to much higher $J$ levels.  Thus, observations of
the CO ladder allow us to discriminate between the different types of
radiation fields.

Recently, it has become clear that PDR and XDR excitation is not
sufficient to fully explain observed molecular line emission ratios.
For instance, \cite{loenen2008} have found that the HCN, HNC, and
HCO$^+$ line ratios observed towards (ultra-)luminous infrared
galaxies span a parameter space that cannot be reproduced by models in
which FUV or X-ray radiation dominates. In their study they showed
that an additional heating mechanism is required, which they suggest
to be mechanical heating caused by dissipating supernova shocks,
injected on large scales and cascading down through turbulent
dissipation to the smallest scales.  More recently,
\cite{papadopoulos10} has suggested that cosmic ray ionization rates in such
galaxies may exceed those measured in the Milky Way by factors of
$10^3$ to $10^4$, hence significantly affect the ionization balance
and thermal-chemical structure of interstellar clouds.  The effects of
high cosmic ray rates on ISM chemistry is studied in detail by
\cite{meijerink11} (who consider mechanical heating as well only for
selected models) and \cite{bayet11-1}.

However, neither the models by \cite{loenen2008} nor those by
\cite{meijerink11} treat mechanical heating self-consistently, as 
it is only implemented as an additional heating term and not as
feedback from hydrodynamics.  The dynamical evolution of the
interstellar medium in galaxies in general and close to a
super-massive black hole in particular, has been studied in 3D
hydrodynamic simulations \citep{wada02, wada05}
but these in turn lacked chemistry, and line emission maps of the
various molecules were created by using constant abundances.  A study
by \cite{perez11} improved on this by using chemical
abundances obtained from the \cite{meijerink2005-1} XDR code.
Nevertheless, these simulations still lack UV input and feedback from
the hydrodynamics on the chemistry still needs to be done
self-consistently.

The dominant coolants in the ISM which are commonly observed are
the [CI] 609 $\mu m$, [CII] 158 $\mu m$ and the [OI] 63 $\mu m$
lines. These fine-structure lines are not sensitive to additional
mechanical heating, when the clouds are already illuminated by FUV
radiation \cite{meijerink11}. This is in contrast to the molecular
species that are formed in regions of the cloud, that are shielded
from irradiation. Thus in this paper we consider molecular species
as possible candidates for tracing regions which are dominated by
mechanical heating. Although a simple dust model is 
taken into account \citep{meijerink2005-1}, dust diagnostics are not discussed 
since these diagnostics are out of the scope of this paper.  We will explore the
possible tracers in a number of steps.  In this paper, we will first
establish under which physical conditions mechanical heating is
important. We will estimate what range of mechanical heating rates we
may expect for the different densities in the ISM. We will then
determine how this affects the thermo-chemical balance in the models,
and, by way of first application, discuss the effect on the integrated
column densities of common molecular species such as CO, H$_2$O, HNC, HCN and HCO$^+$,
and some of their ratios. In subsequent papers, we will calculate line emission maps
that can be compared directly to the detailed observations obtained
with (sub)millimeter telescopes, notably the Atacama Large Millimeter
Array. Ultimately, our aim is to couple chemistry and hydrodynamics in
order to treat the relevant physical effects in a self-consistent
manner.

\section{Methods \label{sec:methods}}

We solve the equilibrium state of all model clouds by using an
optimized version\footnote{The optimized code is called PDR-XDR-1.1} of the
Leiden PDR-XDR code developed by \cite{meijerink2005-1} (Details of the optimization
and further improvements are mentioned in Sect.\,\ref{subsec:numericalSchemes}). In 
its present form, the code assumes a 1D PDR geometry where various heating and cooling processes
are included in solving for the thermal and chemical balance. The
radiative transfer for the cooling lines and the escape probabilities
are computed based on the work by \cite{deJong80}.  We refer the
reader to \cite{meijerink2005-1} for further details on the
thermal processes and chemical network used in solving for the steady
state equilibria.

\subsection{Parameter space and reference models}

In order to reduce the number of free parameters, we considered three
metallicity levels, the benchmark being $Z = Z_\odot$ (solar
metallicity) to which we added $Z = 0.5 Z_\odot$ (characteristic of
moderately metal-poor dwarf galaxies) and $Z = 2 Z_\odot$
(characteristic of spiral galaxy center conditions).  Throughout the
paper, our discussion concentrates on the $Z = Z_\odot$
models\footnote{The analogous grids for the other two metallicities
  are presented in the Appendix}.  For the cosmic-ray ionization rate,
we adopted the value $5 \times 10^{-17}$s$^{-1}$ used in \cite{meijerink2005-1}
 which is close to that in the solar neighborhood.  The grids consisted of 
models spanning a wide range of
physical conditions, uniformly sampled (in log$_{10}$ scale) in ambient gas
density $1 < n < 10^6$\,\cmt, incident radiation field $ 0.32 < G_0 <
10^6$ and mechanical heating $ 10^{-24} < \Gamma_{\rm mech} <
10^{-16}$\,\ecs.  In the absence of mechanical heating, $n$ and $G_0$
dominate the characteristics of a PDR. The ambient gas density $n$ is
the total number density of hydrogen nuclei, which is set constant
throughout the slab. The values considered cover the full range of
densities encountered in the diffuse interstellar medium to those of
dense molecular cloud cores. The incident FUV radiation field $G_0$ is
measured in so-called Habing units \citep{habing1969BAN} corresponding
to $1.6\times10^{-3}$\,erg cm$^{-2}$ s$^{-1}$.  The values considered here
include radiation fields ranging from only a few times stronger than
the relatively low solar neighborhood field to the intense fields
irradiating clouds close to young OB associations; the upper limit
corresponds to the radiation field intensity at a distance of about
0.1 pc from an O star. 

In constraining a physically relevant range of mechanical heating
  rates, we took a combination of values recovered from softened
  particle hydrodynamics simulations (SPH hereafter) \citep{intiPhdT},
  and added estimates of mechanical heating rates induced by
  turbulence caused by supernova shocks that are dissipated in the ISM
  \citep{loenen2008}. The SPH simulations are of dwarf galaxies with
  $n$ up to $\sim 10^3$\,\cmt~ where the mechanical heating rates
  ranged from $\sim 10^{-30}$ to $10^{-22}$\,\ecs~for such
  densities. When naively extrapolated to densities of $10^6$\,\cmt,
  rates up to $10^{-18}$\,\ecs~are expected. In the \cite{loenen2008}
  recipe, it is assumed that $\eta = 10\%$ of the supernova blast
  energy ($E_0 = 10^{51}$\,erg per SN event) is absorbed throughout
  the continuous starburst region (a region which by definition 
  sustains a considerably high SFR, which is steady, by consuming the
  available gas), whose size is taken to be $D_{\rm SB} =
  100$\,pc. Eq.~\ref{eq:snr}, which is a combination of Eq.~3 and 4 in
  \cite{loenen2008}, relates $\Gamma_{\rm mech}$ to the the supernova
  rate (SNR).

\begin{equation}
  \Gamma_{\rm mech} = \frac{ {\rm SNR} E_0 \eta} { V_{\rm SB} V_{\rm PDR} n_{\rm PDR} } \label{eq:snr}
\end{equation}

\noindent where $V_{\rm SB}$ is the volume of the starburst region, whereas $V_{\rm PDR}$ and $n_{\rm PDR}$
are the volume and number density of PDRs in that starburst region respectively.  They also assumed that the absorbed 
energy is distributed evenly among the PDRs in this starburst region. The 
number of the PDRs in the starburst region is estimated by assuming a gas density contrast 
of 10 (the inverse of the filling factor) between that of the PDRs and the ambient ISM of 
the starburst.  Considering a Salpeter IMF for the stellar population, the star formation rate 
(SFR) is related to the SNR via ${\rm SNR} / {\rm SFR} = 0.0064 $ (see \cite{dahlen99} for details).
In computing the SNR, they assumed that stars with $M > 8$~M$_\odot$~end up as supernovae (see 
\cite{loenen2008} for more details on the estimates and assumptions). A \cite{kroupa2001} IMF, over
the mass range from $0.1 M_\odot$ to $125 M_\odot$ considered, would result in a marginally higher ${\rm SNR} / {\rm SFR} \sim 0.01 $.
This higher rate reflects a slightly enhanced mechanical heating rate. However, the difference in the resulting $\Gamma_{\rm mech}$ between 
both IMFs over the same mass range is $\sim 40\%$, thus for simplicity we adopted the Salpeter IMF.

By way of example, a mechanical heating rate $\Gamma_{\rm mech}$ of $2
\times 10^{-20}$~\ecs~is expected in a quiescent disc, such as that of
the Milky Way with an SFR $\sim 1$~M$_\odot$~yr$^{-1}$ and an SNR of
$\sim$0.01~yr$^{-1}$~. In regions with active star formation
(SFR$\sim 50$~M$_\odot$~yr$^{-1}$ and SNR$\sim$0.3~yr$^{-1}$) the
mechanical heating rate is much higher, $\Gamma_{\rm mech} = 1 \times
10^{-18}$~\ecs~. For extreme starbursts with (SFR$\sim 1000$~M$_\odot$~yr$^{-1}$ and
SNR$\sim$6.4~yr$^{-1}$) mechanical heating rates $\Gamma_{\rm mech} =
2 \times 10^{-17}$~\ecs~are possible.  Another source of mechanical energy input is the
  outflows of young stellar objects (YSOs). The outflow phase is
  short-lived, and as discussed by \citep{loenen2008} a nearby dense
  cloud ( $n > 10^5$~\cmt) is required for a significant amount of
  energy to be absorbed by the surrounding ISM (compared to the
  mechanical heating due to supernovae). Hence energy input due to
  YSOs is ignored in this study. Based on a similar order of
  magnitude estimate, mechanical energy input due to stellar winds are
  also ignored, since they would contribute up to 6\% to the total 
  mechanical heating \citep{intiPhdT, sb99}.

It is of interest to determine the effects of mechanical heating by
comparing the results obtained from models with and without a
mechanical heating term.  To this end, we have selected from the grids
calculated for analysis and discussion in this paper the same
reference models that were studied in detail by \cite{meijerink2005-1}
without taking into account mechanical heating.  These models are
summarized in Table\,\ref{tbl:refModels}.

\begin{table}[h]  
\centering
\begin{tabular}{c c c}
  \hline
  Model Name & $\log n$ (cm$^{-3}) $ & $\log G_0$\\
  \hline
  \hline
  M1 & 3.0 & 3.0 \\
  M2 & 3.0 & 5.0 \\
  M3 & 5.5 & 3.0 \\
  M4 & 5.5 & 5.0 \\
  \hline
\end{tabular}
\caption{Parameters of the reference models. These are identical to
  the ones in \cite{meijerink2005-1}. M1 and M2 correspond to low
  density clouds in galactic centers in the presence of a starburst
  event, whereas M3 and M4 correspond to much denser clouds where
  excitation of high density gas tracers such as HCN is possible.
\label{tbl:refModels}}
\end{table}

In this first exploration, $\Gamma_{\rm mech}$ is added per unit
volume. In doing so for a certain grid, each model in the $n - G_0$
parameter space has an extra $\Gamma_{\rm mech}$ added to its
heating budget. Another choice could have been the addition of
$\Gamma_{\rm mech}$ per unit mass, where for a certain grid the
amount of added mechanical heating would be proportional to the gas
density of the model in the grid. For simplicity we considered
adding mechanical heating per unit volume.

\subsection{Numerical schemes \label{subsec:numericalSchemes}} 

Finding the thermal and chemical equilibrium for each slab is an
$(N+1)$-dimensional numerical root finding problem. $N$ being the
number of species (atomic and molecular species and electrons) in the 
chemical network in addition to the thermal balance. It involves 
solving the set of non-linear equations Eqs.\,\ref{eq:specRates} 
and \ref{eq:thermal}.

%----------
\begin{equation}
  d\mathbf{n}(T)/dt = \mathbf{0} \label{eq:specRates}
\end{equation}
%----------

%----------
\begin{equation}
  \Gamma(T,\mathbf{n}) - \Lambda(T,\mathbf{n},t) = 0\label{eq:thermal}
\end{equation}
%----------

\noindent The first equation (chemical balance) is the rate of change
of the density of the species at a certain temperature $T$. We solved
for this vector equation of length $N$ following the recipe by
\cite{tielens1982} in using the multi-dimensional Newton-Raphson
method.  The thermal balance is solved for in the second equation,  where
$\Gamma$ and $\Lambda$ are the total heating and cooling rates.  These two terms 
are a 1D function of $\mathbf{n}$ (used as input from the
previous equation) and $T$. This process is repeated iteratively until
convergence with a tolerance of $10^{-6}$ for the chemical balance and
$10^{-3}$ for the thermal balance.  Two major improvements lead
to a speed-up by a factor of $\sim 30$ over the original PDR-XDR code.
These are (a) tweaking the LU decomposition (optimizing compiler
flags and matrix elements storage scheme) used in the multi-dimensional
Newton-Raphson root finding, and (b) using in the thermal balance root finding the
secant method (with a convergence order of $\sim 1.6$) instead of the bisection method
which has a linear convergence rate.

The distance from the surface of the cloud is measured in terms of the
visual extinction ($A_V$) due to interstellar dust. It is related to
the total column density of hydrogen nuclei $N(\rm H)$ according to
Eq.\,\ref{eq:Av} \citep{bohlin78}.

%-------------
\begin{equation} \label{eq:Av}
  A_V = 5.34 \times 10^{-22}\,N_{\rm H}\,Z \textrm{ mag cm}^{-2}
\end{equation}
%-------------

\noindent Once the equilibrium state of the surface slab (with zero
thickness) is solved for, the width of the remaining slabs is chosen
adaptively while resolving the transition zone which has large temperature
gradients compared to the small temperature gradients at the surface of the slab
and at high depths. This is also one of the changes over the original implementation. In choosing 
the slab width adaptively, the maximum allowed relative difference in the
temperature between consecutive slabs was set to $5\%$. Smaller values were
tested as well, where no significant difference was noticed.

It is assumed that the mechanical heating due to turbulence is absorbed by the
ISM at all scales. Hence for simplicity $\Gamma_{\rm mech}$ is added \emph{uniformly}
throughout the cloud.

Molecular clouds may have a visual extinction up to 200 mag
\citep{tielens1982}. Beyond $A_V \sim 5$\,mag most of the species become
molecular whereas for $A_V > 10$\,mag all abundances are almost constant
since physical conditions do not change anymore.  The maximum depth, we
allowed for, in all the models, was $A_V=20$\,mag.  This corresponds
to $N({\rm H}) = 7.5 \times 10^{22}$, $3.7 \times 10^{22}$ and $1.8 \times 10^{22}$\,\cms\,
for metallicities of $Z=$0.5, 1 and 2$Z_\odot$ respectively. This enables us to
 compare clouds of different metallicities up to the same column density.  In doing so, 
 a fixed $N({\rm H}) = 1.8 \times 10^{22}$\,\cms~ (the maximum $N({\rm H})$ for $Z=$2$Z_\odot$) was adopted
in the illustrations where column densities of molecular species were used.

For each value of mechanical heating and metallicity, $25 \times 25$
models equally spaced in $\log_{10} n$ and $\log_{10} G_0$ were
calculated, up to a total of $\sim $16000 models.  When there is no or
very little mechanical heating, the temperature may drop below 10~K at
a certain depth in the cloud.  Since the code is designed to work for 
$ 10 < T < 10^4$~K$^[$\footnote{The lower bound in the temperature is 
    set to 10~K since many reaction constants become inaccurate
    below that value \citep{rolling11}. He and Fe$^+$ cooling become
    important above $10^4$~K which make the thermal balance inaccurate since
    these are not included in the current version (in addition to other 
    dielectronic recombination processes). But below $10^4$~K, the contribution of
    these processes is negligible to the total cooling. Also for $T > 10^4$~K, the heating
    efficiency expression becomes inaccurate. However, not many models exceed $10^4$~K, 
    but we allowed that since it is needed to make the models converge as
    the solution is progressed into the cloud.}$^]$, for slab elements where the
lower limit is reached we solved the chemical network for a fixed temperature of 10~K.
In making the convergence more robust,
we built a small database of guess values (used as input in the
root finding) for the surface slab. This enabled us to explore regions
in the parameters where the original implementation failed.

In running the models, an improved version of the PDR-XDR code was
incorporated into the Astrophysical Multi-purpose Software Environment
(AMUSE) framework. The AMUSE package allows astrophysical codes from
different domains to be combined to conduct numerical experiments (see
\cite{pelupessy12} for a more complete description). It is a
development of the Multi-physics and Multi-scale Software Environment
\citep[MUSE, ][]{PortegiesZwart2009} and is freely available for
download\footnote{\texttt{www.amusecode.org}}.  (The PDR code itself
will be made available in the future). The interface to the Meijerink
PDR-XDR code takes the main parameters $G_0$, $n$, $\Gamma_{\rm mech}$ as
input and calculates the PDR equilibrium properties.  This will allow
the code to eventually be used as an equilibrium sub-grid model for
e.g. galaxy scaled hydrodynamic simulations. In the present work this
feature is not used, only the capability to run many models in
parallel is.

\section{Results}

In this section we present the comparisons for thermal and chemical
diagnostics of the grids modeled with and without mechanical
heating. We start by discussing the temperatures at the cloud surface,
and at the maximum cloud depth considered since for a given model, the
gas temperature to a large extent determines the chemical composition.
Following this, we will discuss the statistical properties of the CO, H$_2$O,
HCN, HNC and HCO$^+$ molecules which are important diagnostics for
moderate and dense gas. Throughout this section, we use the reference
models (M1 to M4) to illustrate the impact of mechanical heating on
the properties of the PDRs as a function of depth.

\subsection{Gas temperature}

In the absence of other heating terms (such as due to mechanical
heating) PDRs are characterized by rapid decreases in temperature,
from $\sim$1000~K at the surface down to $\sim$10~K in regions where
$A_V$ exceeds 10 mag.  Since mechanical heating may be expected to
considerably change the temperature structure, we will first explore
this. 

\subsubsection{Temperature at the surface \label{subsec:surfTemp}}

\begin{figure*}[!tbh]
  \centering
  \includegraphics[scale=1.5]{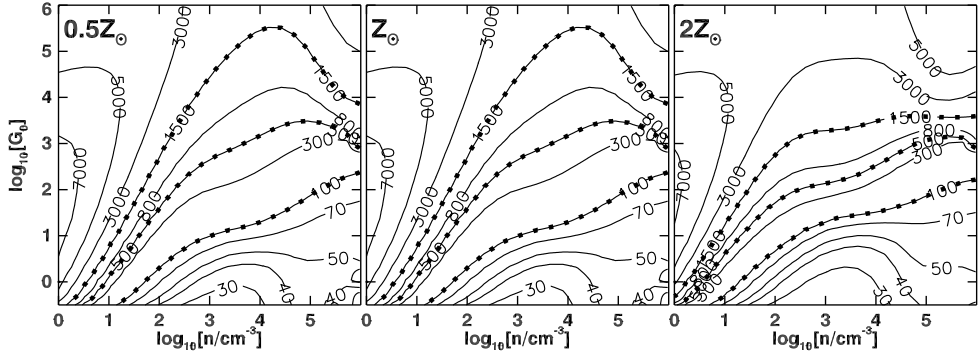}
  \caption{Temperature profile (without mechanical heating) at the cloud surface for
    different metallicities.  Generally, the temperature rises as a
    function of $G_0$ and constant $n$, since the dominant heating
    term is the photo-electric heating, which is proportional to $G_0$.
    However the behavior for constant $G_0$ as a function of $n$ is not as
    trivial. See Section 3.1 of \cite{kaufmann99} and Section 3 of
    \cite{meijerink2007-1}~ for more discussion of the topology of
    the contours.  This version of the code covers a larger parameter
      range than \cite{meijerink2007-1}~. Solid lines with filled squares
      highlight the temperature contours for 100, 500 and 1500~K~. The most
      significant change when comparing the three panels is for the 1500~K~ contour,
      where for $Z=2Z_\odot$ higher temperatures are attained ( compared to the 
      lower metallicities) in the same part of the parameter space ( $n > 10^3$~\cmt~, 
      $G_0 > 10^4$~) .
    \label{fig:surfaceTemp} }
\end{figure*}

At the surface we only need to solve for the equilibrium of a single
slab (with zero thickness).  Because this can be done much faster
than in the case of a full slab, a higher resolution is possible for
the surface grids.  These surface grids therefore consisted of 2500
(50 $\times$ 50) models uniformly sampled in $\log_{10} n$ and
$\log_{10} G_0$. In Fig.\,\ref{fig:surfaceTemp} the surface temperature,
without mechanical heating ($\Gamma_{\rm mech} = 0$\,\ecs), is shown for three
different metallicities. The temperature contours are in good
agreement with the profiles recovered by \cite{kaufmann99} and 
\cite{meijerink2007-1}(which uses the original implementation). We note 
that metallicity has little effect on the overall topology of the contour 
plots, except for regions where
$G_0 > 10^4$, especially where $n > 10^4$\,\cmt. The temperature almost
doubles as $Z$ is increased from $Z_\odot$ to $2Z_\odot$, with no
noticeable change between $Z=0.5Z_\odot$ and $Z=Z_\odot$. This is
because cooling scales as $Z$ whereas the dominant heating term scales
almost as $Z$ for low $n$ but increases to $\sim Z^2$ as $n$ increases
to $10^6$\,\cmt. This temperature dependence on metallicity has also
been observed by \cite{rolling06}. It is worth noting that in the
lower right-hand corner of the grid ($n > 10^{5.5}$\,\cmt, $G_0 < 5$)
three equilibria were found, one unstable and two stable ones. The
stable equilibria correspond to a low ($\sim 50$~K) and a high
temperature ($\sim 300$~K) state with the unstable one in-between. Here
we have shown the low-temperature equilibrium. In any case, we note
that this part of the parameter space is not of physical relevance for
systems that interest us here..

%%% 3
The dominant heating term at the cloud surface is 
photo-electric heating, where electrons ejected from the dust grains
collide elastically with the gas and heat it.  The typical efficiency
($\epsilon$) of this process is quite low 0.5 to 1\% with a maximum of $\sim 5\%$
\citep{1977ApJS...34..405B}. The photo-electric heating rate
($\Gamma_{\rm photo}$) at the surface is given by the expression
\citep{bakesTeilens94}:

\begin{equation} \label{eq:gammaPhoto}
  \Gamma_{\rm photo} = 10^{-24} \epsilon G_0 n \textrm{ erg cm}^{-3} \textrm{s}^{-1}
\end{equation}

\noindent where $\epsilon$ depends on electron density ($n_e$), $G_0$ and gas
temperature. The expression for $\epsilon$ is given in Eq-\ref{eq:epsilon} \citep{bakesTeilens94}.

\begin{equation} \label{eq:epsilon}
  \epsilon = \frac{ 4.87 \times 10^{-2}}{1 + 4 \times 10^{-3} \left( G_0 T^{1/2} / n_e \right)^{0.73}} + 
             \frac{ 3.65 \times 10^{-2} \left(T / 10^4 \right)^{0.7} }{ 1 + 2 \times 10^{-4} \left( G_0 T^{1/2} / n_e \right)}
\end{equation}

\noindent Hence
 the $log_{10}$ of Eq.~\ref{eq:gammaPhoto} is linear in 
$\log_{10} n$ and $\log_{10} G_0$  whenever $\epsilon$ is constant.  
Fig.\,\ref{fig:surfGammaPhotoNoMech} shows the total
heating in the $n - G_0$ plane. In the top part of the plot, the
contours are almost straight lines. This is not surprising, $\Gamma_{\rm photo}$
accounts in most cases for more than $95\%$ of the total
heating budget.  To illustrate this, we look at the logarithm of
Eq.\,\ref{eq:gammaPhoto}, which in expanded form becomes :

%-----------
\begin{figure}[!tbh]
  \centering
  \includegraphics[scale=2.0]{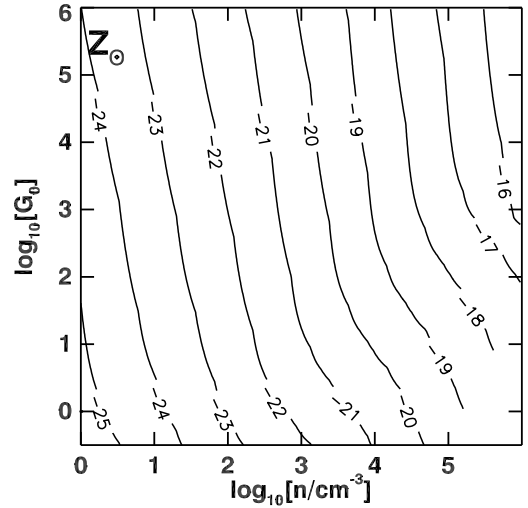}
  \caption{Total heating at the cloud surface without mechanical
    heating. The main characteristic of this plot is that above the
    diagonal, the contours are parallel straight lines; whereas below
    the diagonal $( G_0 = n$ line), there is a break in the slope, indicating that
    photo-electric heating is no longer the dominant mechanism in that
    region. The minimum total heating is of the order of $10^{-25.5}$\,\ecs~
    at the lower left corner and $10^{-15.2}$\,\ecs~at the top right
    corner. \label{fig:surfGammaPhotoNoMech}}
\end{figure}
%-----------

\begin{equation} \label{eq:gammaPhotoLog}
  \log_{10} \Gamma_{\rm photo} = -24 +  \log_{10} \epsilon + \log_{10}G_0 + \log_{10} n 
\end{equation}

\noindent which is linear in both $\log_{10}G_0$ and $\log_{10}
n$. This means that the heating efficiency is almost constant wherever
the contours are straight lines. This is not true in the regions below
the break in the contour lines which occurs in the region under the
  $G_0 = n$ line in Fig.\,\ref{fig:surfGammaPhotoNoMech}. In that part of parameter space,
photo-electric heating is complemented by H$_2$ photo-dissociation
heating.  However, this is a small part of the total parameter space,
so that we are justified in using photo-electric heating as a good
approximation for the total heating.  Hence, we expect total heating
to have a linear dependence on $\log_{10} n$ for constant values of
$G_0$. This is achieved by fitting horizontal cuts of $\Gamma_{\rm total}$
in Fig.\,\ref{fig:surfGammaPhotoNoMech} as a function of $n$ for
different values of $G_0$.

From the fits, we could see that the minimum surface heating for the
lowest density gas considered ($n=1$~\cmt) ranges from
$10^{-25.6}$\,\ecs~(for low $G_0$) to $10^{-24.0}$\,\ecs~(for high
$G_0$).  In contrast, the maximum surface heating for $n = 10^6$\,\cmt,
is $10^{-15.2}$\,\ecs.  We also observe that the surface heating scales as
$n^{1.21}$ for $G_0 < 10^2$ and $n^{1.55}$ for $G_0 > 10^4$.
%%% 4

The equilibrium temperature is expected to vary depending on the
amount of extra heating introduced via $\Gamma_{\rm mech}$. Following this
simple assumption, we expect to have significant changes in the
equilibrium temperature when $\Gamma_{\rm mech}$ is comparable to the
total surface heating, with a varying impact on different parts of the
parameter space.  In order to determine the zone in the parameter
space where $\Gamma_{\rm mech}$ alters the equilibrium state, we solve for
$n$ in Eq.\,\ref{eq:gammaPhotoLog} by equating it to $\Gamma_{\rm mech}$. In
other words, we solve for $n$ in $\Gamma_{\rm total,surface}(n) = \Gamma_{\rm mech}$
for the range in $\Gamma_{\rm mech}$ we have considered in our parameter
space. These solutions, $n_c$, listed in
Table\,\ref{tbl:criticalDensities} mark the lines in the $G_0 - n$
plane where mechanical heating is the same as the total surface
heating. To the left of these lines, $\Gamma_{\rm mech} >
\Gamma_{\rm surface}$.

%-------------------------
%-------------------------
%-------------------------
\begin{table}[h]
\centering
\begin{tabular}{|c|c|c|c|c|c|c|c|c|c|}
  \hline
  $\log_{10} [\Gamma_{\rm mech} /$\ecs$]$ &  -24  & -22  & -20  & -18  & -16  \\
  \hline
  $\log_{10} [n_c/$\cmt$]$               &  0.6 &  1.9 &  3.2 &  4.5 &  5.9 \\
  \hline
\end{tabular}
\caption{Estimated values of gas density, where clouds with lower densities (listed in
    the bottom row), the shape and equilibrium temperature of the contours of Fig.\,\ref{fig:surfaceTemp}
  are significantly effected by $\Gamma_{\rm mech}$ (top row).~ For $\Gamma_{\rm mech} < 10^{-18}$\,\ecs, 
  the fitting parameters for $\log_{10} G_0 = 2$ are used. For higher mechanical heating rates, those
  for $\log_{10} G_0 = 5$ are used since in that region of the parameter space $\Gamma_{\rm mech}$ is 
  more relevant to high $G_0$. \label{tbl:criticalDensities}}
\end{table}
%%% 5 
%%% 6

\begin{figure*}[!tbh]
  \centering
  \includegraphics[scale=1.5]{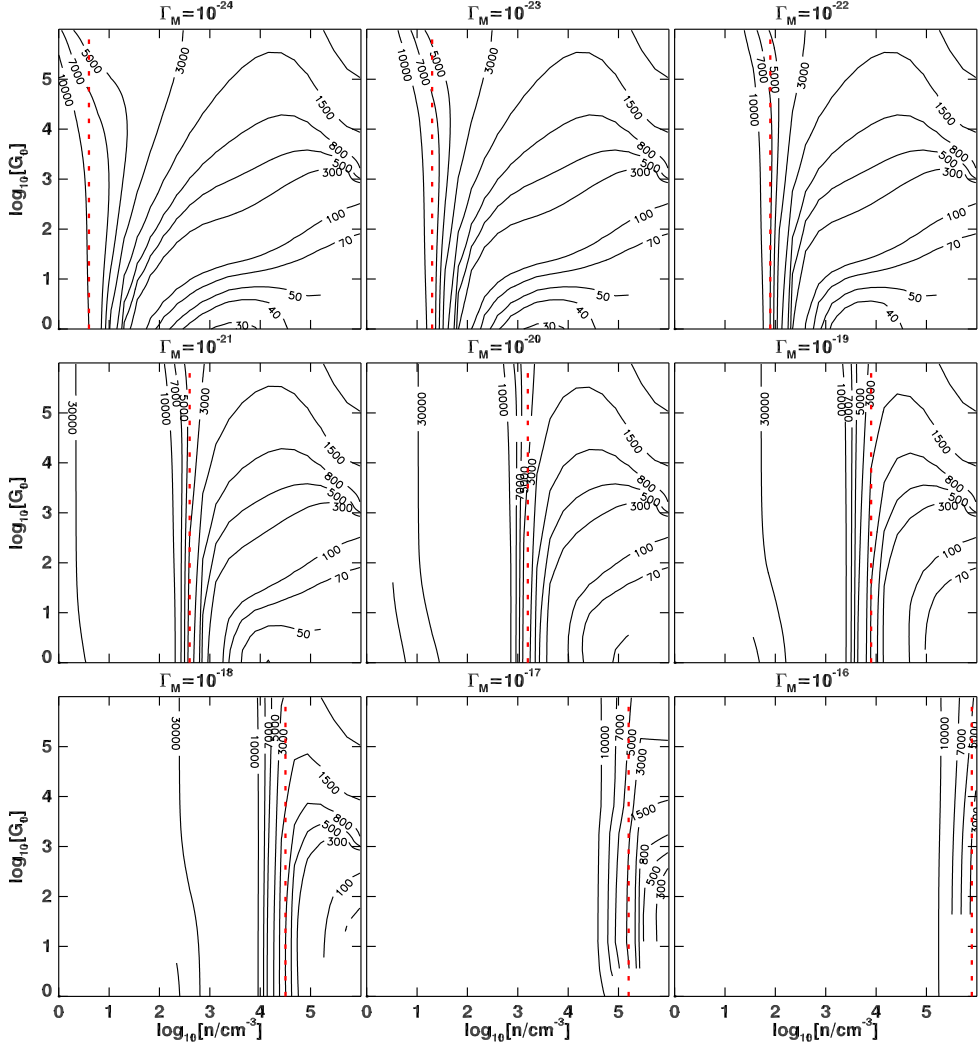}
  \caption{Surface temperature (for $Z = Z_\odot$) for different values of mechanical
    heating. Each panel shows the surface temperature contours for a different
    value of $\Gamma_{\rm mech}$ mentioned at the top of each panel.  In probing
    the effect of mechanical heating, each panel should be compared to the middle
    panel of Fig.\,\ref{fig:surfaceTemp} which is also for $Z = Z_\odot$. Different
    parts of the parameter space in $n$ and $G_0$ are affected differently, depending
    mainly on the gas density. For $\Gamma_{\rm mech} < 10^{-21}$~\ecs~, regions in
    density with $n < 10^3$~\cmt have their temperatures increase significantly. 
    The vertical dashed red lines mark densities $n = n_c$ where 
    $\Gamma_{\rm mech} = \Gamma_{\rm total,surface}$. The location of these lines moves
    from very low densities ( $n \sim 10^0.5$~\cmt ) to very high densities ( $n > 10^5$~\cmt )
    as $\Gamma_{\rm mech}$ increases from $10^{-24}$ to $10^{-16}$~\ecs, sweeping the 
    parameter space in density (for a certain grid) from left to right. In general the 
    contour lines at $n = n_c$ become vertical, indicating that the surface temperature
    becomes independent of $G_0$. In subsequent plot, those lines are plotted only for $Z = Z_\odot$ 
    grids, since the fits from which $n = n_c$ were derived were done only for solar metallicity.
  \label{fig:surfTempMech}}
\end{figure*}

  \noindent In Fig.\,\ref{fig:surfTempMech} the equilibrium
  temperature of the surface slab is shown for the range of
  $\Gamma_{\rm mech}$~ that we consider. Regions, that are first
  affected by the mechanical heating input, are the ones where
  $\Gamma_{\rm mech}$~ is comparable to the surface heating without
  mechanical input (see Fig-\ref{fig:surfGammaPhotoNoMech}).  This
  behavior can be traced by following the location of the dashed
  vertical lines (where $\Gamma_{\rm mech} = \Gamma_{\rm
    total,surface}$) and comparing the values of $\Gamma_{\rm mech}$
  to the total surface heating contours in
  Fig.\,\ref{fig:surfTempMech}. It is clear that whenever the value of
  $\Gamma_{\rm mech}$ is comparable to the total surface heating, the
  temperature contours become almost vertical in the $n - G_0$ plane
  in Fig.\,\ref{fig:surfTempMech}. The locations where $\Gamma_{\rm
    mech} = \Gamma_{\rm total,surface}$ are marked by the dashed
  vertical lines where $n = n_c$ ( see
  Table\,\ref{tbl:criticalDensities}).  For instance, in the first
  panel in Fig.\,\ref{fig:surfTempMech} ($\Gamma_{\rm mech} =
  10^{-24}$~\ecs), regions where temperatures increase significantly
  are when $\Gamma_{\rm total,surface} < 10^{-24}$~\ecs (see
  Fig.\,\ref{fig:surfTempMech}, and compare first panel in
  Fig.\,\ref{fig:surfTempMech} to the middle panel in
  Fig-\ref{fig:surfaceTemp}). Also we notice that the temperature
  increases from $\sim 3000$~K (for $G_0 \sim 10^{2}$) to $\sim 10000$~K
  at the density $n = n_c$, whereas it increases from $\sim 5000$~K (for $G_0 \sim
  10^{5.0}$) to $\sim 7000$~K at the same density. This can be
  explained by the fact that heating at the surface is mainly due to
  photo-electric heating, thus as $G_0$ increases (for a fixed $n$),
  the amount total heating increases. Hence a fixed amount of
  mechanical heating would contribute less to the total heating
  budget, leading to a lower increase in the surface
  temperature. Subsequent panels in Fig.\,\ref{fig:surfTempMech} which
  correspond to higher amounts of $\Gamma_{\rm mech}$ can be
  interpreted in the same fashion. The main feature of these panels is
  the shift in the position of the line marking $n = n_c$ towards
  higher densities. In the last panel where $\Gamma_{\rm mech} =
  10^{-16}$~\ecs~, the whole parameter space is dominated by
  mechanical heating.

  Regions with a SFR of ~$\sim 1 $~\modotyr are dominated by
  mechanical heating up to densities of $\sim 10^3$~\cmt~. In Extreme
  starburst, where SFR of $\sim 1000 $~\modotyr are possible, all
  clouds with $n < 10^{5.5}$~\cmt~ become dominated by mechanical
  heating. In both cases, surface temperatures of $\gtrsim 5000$~K are
  reached, and the temperature contours become independent of FUV
  heating.

\subsubsection{Temperatures deep into the cloud}

Now we turn our attention to the molecular region. The solution was
terminated well beyond $A_V \sim 1$\,mag (where mainly atomic and radicals
exist)  at $A_V = 20$\,mag (where everything is molecular). As the
FUV radiation penetrates the cloud, it is attenuated by dust
absorption.  Thus, the photo-electric heating that was the dominant at
the surface, rapidly becomes less important and almost vanishes
because (a) $\Gamma_{\rm photo}$ has a $ \exp{(-A_V)}$ dependence on $A_V$
, and (b) the heating efficiency $\epsilon$ becomes very small as
electron abundances decrease significantly by  recombination. 
%%% 7
Thus, in the absence of $\Gamma_{\rm mech}$, heating by cosmic rays and dust (especially
at high $G_0$) become the dominant ones in the molecular region.

\begin{figure}[!tbh]
  \centering
  \includegraphics[scale=2.0]{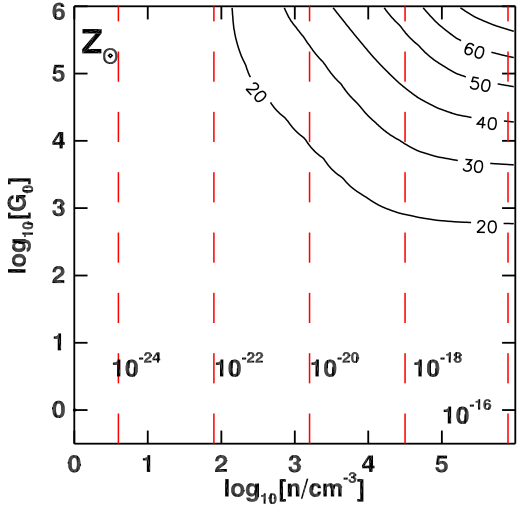}
  \caption{Temperature at $A_V = 20$ without mechanical
    heating. Higher equilibrium temperatures are attained for dense
    molecular clouds (top right corner) with decreasing equilibrium
    temperature towards low $n$ and $G_0$. Below the 20~K contour, all
    temperatures are set to 10~K -- the minimum allowed in the code.
    The red dashed lines mark the densities $n=n_c$ where $\Gamma_{\rm mech} =
    \Gamma_{total,surface}$ from Table\,\ref{tbl:criticalDensities}.
    \label{fig:insideTempMechNoMech}}
\end{figure}

In Fig.\,\ref{fig:insideTempMechNoMech} we show contours of the
equilibrium temperature at $A_V = 20$.  In the absence of any
mechanical heating, and in contrast to the situation for the surface
slab, the topology of the contours is quite simple.  The highest
temperatures, with a maximum of $\sim 80$~K, are in the upper right
corner at $n = 10^6$\,\cmt~and $G_0 = 10^6$.  Temperatures decrease
monotonously along the diagonal until they reach 10~K, which is the
minimum temperature we allowed for.  The main characteristic of a PDR
is that the surface is heated to high temperatures, which is a thin
layer.  Beyond this layer, in the transition zone, the temperature
drops sharply. Although the temperature profile in
Fig.\,\ref{fig:insideTempMechNoMech} is simple, it is
counter-intuitive.  We might expect the temperature to decrease as the
density increases at a fixed $G_0$, but the opposite occurs.  We
explain this by the fact that as $n$ increases, the coupling between
gas and dust also increases.  Dust temperature depends on $G_0$ and
decreases as a function of $A_V$.  Nevertheless, for very high $G_0 >
10^3$, the decreasing dust temperatures are still high enough to heat
the gas ($\sim 30$~K).  

\begin{figure}[!tbh]
  \centering
  \includegraphics[scale=1.0]{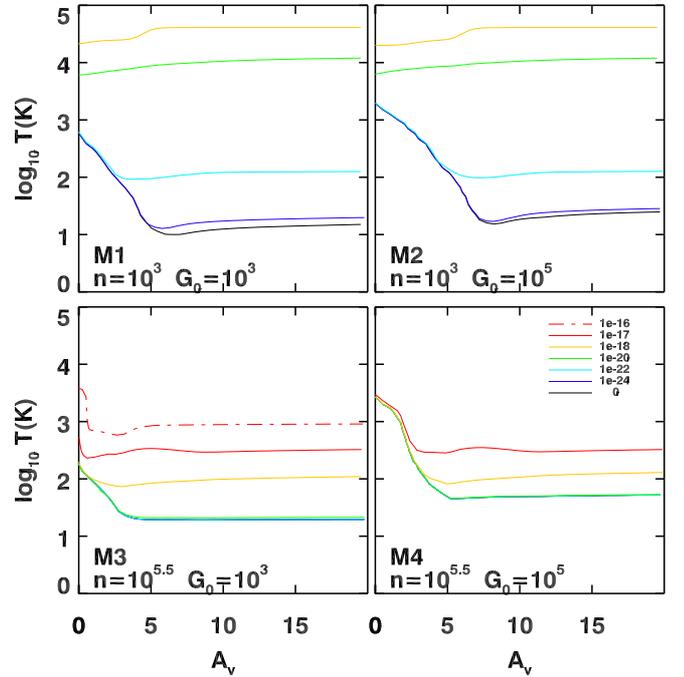}
  \caption{Gas temperature as a function of the visual extinction for
    selected models. In the top two panels, for $\Gamma_{\rm mech} >
    10^{-22}$\,\ecs~(green and orange curves) the transition from H to H$_2$
    does not occur. For the high density models the transition fails
    only for M4 when $\Gamma_{\rm mech} > 10^{-17}$\,\ecs.}  
\label{fig:selectedModelGasTemp}
\end{figure}

  The temperature profile throughout the cloud is shown in
  Fig.\,\ref{fig:selectedModelGasTemp} for the selected models M1 to
  M4. The temperature generally decreases as a function of $A_V$. For
  some models which have high mechanical heating rates, the
  temperature increases and then saturates at higher values than those
  obtained at for the surface. As for the models whose temperature
  decreases, we see that as the density decreases, more specifically
  as the ratio $G_0 / n$ increases, the point of saturation of the
  temperature gets shifted further away from the surface of the cloud
  ($A_V=0$\,mag). For instance, when looking at clouds illuminated
  with a high FUV flux of $G_0 = 10^3$, the saturation point gets
  shifted from $A_V \sim 5$\,mag for $n = 10^3$\,\cmt~ (low density
  gas) to $A_V \sim 2$\,mag for $n = 10^{5.5}$\,\cmt~(high density
  gas, compare left panels of Fig.\,\ref{fig:selectedModelGasTemp}). A
  similar behavior is seen for the right panels which have a higher
  FUV flux with $G_0 = 10^5$.  However, we note that the saturation
  point remains almost constant for a fixed $G_0 / n$, which can be
  seen in comparing M1 and M4, where $G_0 /n$ are of the same order of
  magnitude. The temperature saturation point is also related to the
  H/H$_2$ and C$^+$/C/CO transition zones.  For example, along the
  lines (in $n - G_0$ parameter space) where $G_0/n = $~ 0.003, 1 and
  300, the location of the transition zones in the chemical species
  are almost unchanged.

In a PDR, the total heating at the surface it is dominated by
$\Gamma_{{\rm photo},~A_V=0} \sim \Gamma_{{\rm total},~A_V=0} = \Gamma_{\rm surface}$,
and decreases as a function of increasing $A_V$. Also total heating generally
decreases as a function of $A_V$. In Section. \ref{subsec:surfTemp} it was demonstrated
that mechanical heating increases the surface temperatures significantly whenever $\Gamma_{\rm mech} = \Gamma_{\rm surface}$~.
In such situations, since mechanical heating is added uniformly throughout the PDR, we expect 
it to have a stronger impact (on the physical and chemical properties) at high depths compared
to those at the surface. Hence $\Gamma_{\rm mech}$ should be the most
dominant heating term beyond the surface, even more dominant than heating
due to cosmic rays which attains a maximum of $10^{-21.5}$~\ecs~ for $n = 10^6$~\cmt.  
In Fig.\,\ref{fig:insideTempWithMech} we show the
corresponding parameter grids as a function of $\Gamma_{\rm mech}$; the
thermal profile in these grids has almost no dependence on $G_0$ and
depends solely on $n$.

  In order to illustrate the degree by which $\Gamma_{\rm mech}$
  increases temperatures throughout the cloud, we look at the curves
  in Fig.\,\ref{fig:selectedModelGasTemp} corresponding to
  $\Gamma_{\rm mech} \ne 0$.  We note that for densities
where $\Gamma_{\rm mech} = \Gamma_{\rm surface}$, listed in
Table\,\ref{tbl:criticalDensities}, gas temperatures increase by almost
two orders of magnitude for the low density models M1 and M2, and by
at least one order of magnitude for the high density models M3 and M4.
In fact, it suffices to have a mechanical heating rate 10 to 100
times smaller than the total heating at the cloud surface for gas
temperatures to increase by a factor of $\sim$2 (for high densities) 
and up to a factor of $\sim$10 (for low densities).
Thus even relatively small mechanical heating rates will
substantially modify the physical parameters deduced from a pure PDR
model.  In a continuous starburst with an SFR $=1$~\modotyr~(which corresponds to 
  $\Gamma_{\rm mech} \sim 2 \times 10^{-20}$~\ecs~using Eq.~\ref{eq:snr}),  the mechanical 
luminosity (heating) is about 1\% of the bolometric luminosity \citep{sb99}.
Which implies that the ratio of mechanical luminosity to UV luminosity (where the emission of star
 forming regions peak) is larger than 1\%~. Especially if we consider higher a SFR $\sim 10$\modotyr~
which translates to a $\Gamma_{\rm mech} \sim 2 \times 10^{-19}$~\ecs~. Assuming that half the 
UV radiation is absorbed by the ISM of the galaxy, we would still be in the range where 
$\Gamma_{\rm mech}$ is at least 1\% of the surface heating (which is due to UV radiation), 
thus significantly altering the equilibrium temperatures of PDRs.

%%% 10

\subsection{Molecular gas tracers}

Now that we have determined the degree to which mechanical heating
alters the thermal properties of PDRs, we examine its effect on
the chemical properties.  Molecular clouds consist mainly of molecular
hydrogen which is not directly observable except under very special
circumstances.  Thus, we must depend on the (sub)millimeter line
emission from various transitions of much less abundant tracer
molecules. Each molecular line represents its own critical density and
temperature. 

In the following sections, we will consider the effect of mechanical
heating on a few species commonly used as tracers. These are CO for relatively
low gas densities and HCO$^+$, HCN, and HNC for relatively high
densities, in addition to H$_2$O which has been also recently
observed in very dense gas environments $n \sim 10^6$\,\cmt.

\subsubsection{CO}

Properties of molecular H$_2$ gas are generally studied through
observations of CO, which is the next most abundant species.
The critical density ($n_{cr}$$^[$\footnote{The critical density $n_{cr}$ 
should not be confused with $n_c$, defined earlier, which is the gas density
where $\Gamma_{\rm mech} = \Gamma_{\rm total, surface}$}$^]$) for the ground-state 
$J$=1-0 $^{12}$CO transition is $n_{cr} \sim$ 2000 cm$^{-3}$. This is in the middle section of
range of densities $n$ considered in our grids.  Thus, the response
of CO to mechanical heating may have important consequences for
estimates of the amount of molecular gas and, for instance, may also
explain the high gas temperature of 100-150~K which have been
estimated for molecular gas in the centers of galaxies with densities
of 100-1000 cm$^{-3}$ \citep{israel2009-2}.

\begin{figure}[!tbh]
  \centering
  \includegraphics[scale=1.0]{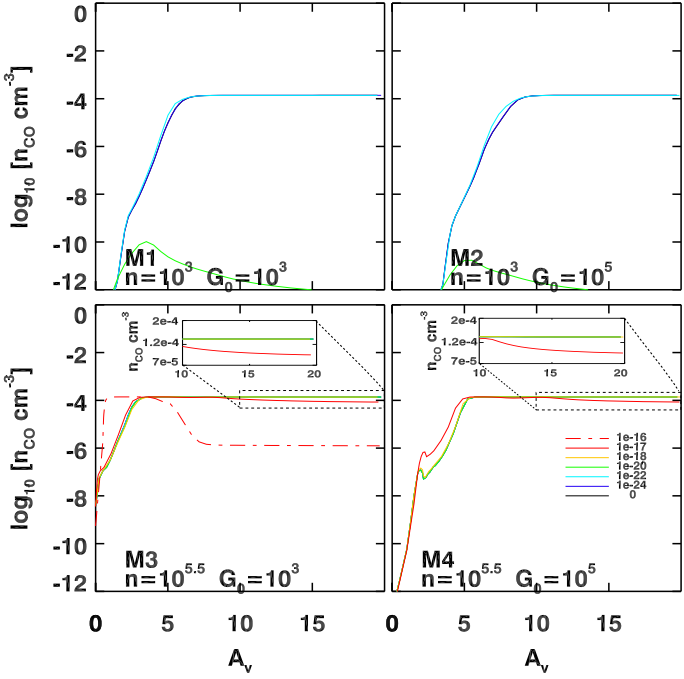}
  \caption{CO abundance as a function of $A_V$ for selected models. In
    all the diagrams, the C/CO transition zone is shifted towards the
    surface as a function of increasing $\Gamma_{\rm mech}$. For M1 and M2, 
    the abundance of CO is below the displayed range for 
    $\Gamma_{\rm mech} > 10^{-20}$\,\ecs\, (these curves are not shown). The other
    curves almost overlap except for $\Gamma_{\rm mech} = 10^{-20}$\,\ecs. In models
    M3 and M4 the abundance of CO drops slightly beyond the transition zone 
    compared to the case of no mechanical heating. The curves overlap except
    for $\Gamma_{\rm mech} = 10^{-16}$~\ecs~where the C$^+$/C/CO transition
    does not fully take place. \label{fig:selectedModelCOAbun}}
  % this is visible in the curves (for M3):
  %                   ../modelData/paper01-models-z1.0/meshes/mesh.dat-id-000029
  %                   ../modelData/paper01-models-z1.0/meshes/mesh.dat-id-000033
  % (for M4)
  %                   ../modelData/paper01-models-z1.0/meshes/mesh.dat-id-000031
  %                   ../modelData/paper01-models-z1.0/meshes/mesh.dat-id-000035
\end{figure}

In Fig.\,\ref{fig:selectedModelCOAbun} we show the abundance of CO as
a function of depth inside the cloud. For the models with densities
close to $n_{cr}$ (M1 and M2), CO abundances decrease sharply for
$\Gamma_{\rm mech} > 10^{-18}$\,\ecs\, and almost disappears when
$\Gamma_{\rm mech} > 10^{-20}$\,\ecs, where $x_{\rm CO} < 10^{-13}$ caused by
very high temperatures $ T > 10000$~K (abundances of species relative to
the total number of hydrogen nuclei are denoted by $x_{i} = n_{i}/(n_{\rm H} + 2n_{{\rm H}_2})$).
At such high temperatures, no molecules, including CO, are formed.
This is due to the very low abundance of H$_2$ which drops from $\sim 0.5$ to $\sim 10^{-9}$
beyond the radical region.  This drop breaks the chemical 
network which leads to the formation of CO. In addition, the higher abundances
of the ionic species H$_3^+$ and He$^+$ enhance the destruction 
of CO through the reactions H$_3^+$ + CO $\rightarrow$ HCO$^+$ + H$_2$ 
and He$^+$ + CO $\rightarrow$ O + C$^+$ + He.  On the other hand,
for $\Gamma_{\rm mech} \le 10^{-18}$ \,\ecs\, the first obvious effect
of mechanical heating on CO is the shift of the C/CO transition 
zone towards the surface of the slab.  This shift is caused by the enhanced
abundance of O$_2$ and OH and higher temperatures in the radical
region (see Fig.\,\ref{fig:selectedModelCOAbun}).

In the high density models M3 and M4, the abundance of CO decreases by
a factor of two as $\Gamma_{\rm mech}$ increases from $10^{-18}$\,\ecs~to 
$10^{-17}$\,\ecs. As $\Gamma_{\rm mech}$ is pushed further to $10^{-16}$\,\ecs~
$x_{CO}$ decreases from $10^{-4}$ to $10^{-16}$~ for M3 and becomes 
under-abundant for M4 ($x_{CO} < 10^{-12}$). This behaviour can be 
explained by the enhanced abundance of the ionised species
He$^+$ and H$_3^+$, which destroy CO through ion-neutral
reactions. Also, the significant drop in abundance of
H and HCO (by three orders of magnitude) reduce the formation
of CO through the reactions H + HCO $\rightarrow$ CO + H$_2$ and O +
C$_2$ $\rightarrow$ CO + C.

\begin{figure}[!tbh]
  \centering
  \includegraphics[scale=2.0]{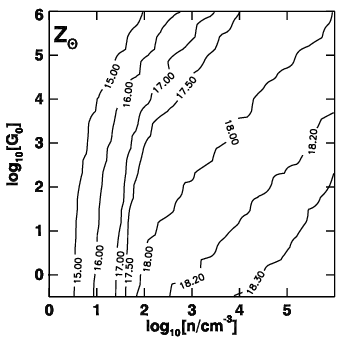}
  \caption{CO total column density up to $N({\rm H}) = 1.8 \times
    10^{22}$~\cms~or $A_V \sim 10$~mag for $Z = Z_\odot$~without
    mechanical heating. \label{fig:COcolDensNoMech}}
\end{figure}

In Figs.\,\ref{fig:COcolDensNoMech} and \ref{fig:COcolDensMech1}, the
column density of CO is shown without and with increasing values of
mechanical heating respectively.  The general trend is an increase
in $N$(CO) whenever the added $\Gamma_{\rm mech}$ is comparable to the surface 
heating of a PDR with no mechanical heating. This occurs at 
densities ($n_c$) marked by vertical dashed lines in 
Fig.\,\ref{fig:COcolDensMech1}. For higher densities in the grids (to the right
of these lines) $\Gamma_{\rm mech} < \Gamma_{\rm surface}$ and the opposite is true 
for lower densities (to the left the lines). At densities a factor of 3 higher than $n_c$ 
(a distance of 0.5, in $\log_{10}$ scale, to the right of $n_c$) where 
$\log_{10} n = \log_{10} n_c + 0.5$ we can see a change in the shape of the contours in $N$(CO).
This change corresponds to an increase
up to a factor of two in $N$(CO), with no noticable change in regions where  
$\log_{10} n > \log_{10} n_c + 0.5$. From our fits of surface heating
these regions have $\Gamma_{\rm surface} < \Gamma_{\rm mech} / 20$. It is
also noticed that the increase of $N$(CO) is more pronounced for high
values of $G_0/n$ as $\Gamma_{\rm mech}$ increases (compare left and right panel 
of Fig.\,\ref{fig:selectedModelCOAbun}). We can explain this by looking at the 
the C$^+$/C/CO transition zone, which gets shifted towards the surface of the
cloud as $\Gamma_{\rm mech}$ dominates. This shift is more pronounced for 
higher $G_0$, thus resulting in more CO in the column of H we are 
considering ($N({\rm H}) = 1.8 \times 10^{22}$\cms).

For conditions similar to that of the Milky Way, SFR $\sim 1$~M$_\odot$~yr$^{-1}$
which corresponds to a $\Gamma_{\rm mech}$ of $2 \times 10^{-20}$~\ecs~, $N$(CO) 
estimates would be marginally ( $\sim 20$\% ) lower when $\Gamma_{\rm mech}$ is 
not taken into account. However, for starbursts with $n \sim 10^{4}$ \,\cmt~ and $G_0 \sim 10^3$
and a significantly higher SFR ($\sim 50$~\modotyr) the $N$(CO) estimates would be 
a factor of two less when the $\Gamma_{\rm mech}$ of $1 \times 10^{-18}$~\ecs~ is ignored
(compare Fig.\,\ref{fig:COcolDensNoMech} to the panel corresponding to 
$\Gamma_{\rm mech} = 10^{-18}$~\ecs~ in Fig.\,\ref{fig:COcolDensMech1}, where the $N$(CO) 
contour is shifted to the right, from $n \sim 10^3$ to $n \sim 10^{4.5}$~\cmt~). For low metallicity grids 
we considered $Z = 0.5 Z_{\odot}$, which may represent systems such as dwarf galaxies 
(where the metallicity may be as low as 0.1 $Z_\odot$),the increase in $N$(CO) may be up to one order of magnitude, reaching $10^{17}$~\cms~.
This occurs even for a low mechanical heating rates corresponding to an SFR of 
$\sim 1$~M$_\odot$~yr$^{-1}$. The increase in $N$(CO) is visible in the 
central part of the parameter space (compare the grids when $\Gamma_{\rm mech} = 10^{-21}$\,\ecs~and
$\Gamma_{\rm mech} = 10^{-20}$\,\ecs~in Fig.\,\ref{fig:COcolDensMech0.5}). In contrast, 
for the highest metallicity we considered, the $Z = 2 Z_{\odot}$ grids, which represents a more 
likely condition for galactic centers, the increase is less prominent even for the 
extreme mechanical heating rates of $\Gamma_{\rm mech} = 10^{-17}$\,\ecs~ corresponding to 
$\sim 500$~M$_\odot$~yr$^{-1}$ (see Fig.\,\ref{fig:COcolDensMech2}). 

Thus, we conclude that we would potentially severely underestimate the
amount of CO in the system if mechanical heating is ignored,
especially for metallicities less than $Z_\odot$, i.e. in dwarf galaxies.

\subsubsection{H$_2$O}
Until the launch of the Herschel Space Observatory, it was not
possible to observe water lines in the far-infrared from
extra-galactic sources, as their emission is blocked by the Earth's
atmosphere. Now, water line emission has been detected by several
Herschel key programs, and a striking example is provided by the water
lines observed in Markarian 231 \citep{gonzales10}. In
this object, water lines intensities are comparable to those of the CO
lines. This is very different from what is observed for, e.g., the
star-burst galaxy M82 \citep{weiss10} and the Orion Bar in the
Milky Way \citep{habart10}.

Water can be formed through a neutral-neutral reaction chain, $\rm O +
H_2 \rightarrow OH + H$, followed by $\rm OH + H_2 \rightarrow H_2O +
H$. In order to obtain significant abundances, the medium has to be
molecular and sufficiently warm, as the reactions have barriers of
$\sim 2000$~K. In X-ray exposed environments, it is possible to obtain
a significant water abundance $x_{\rm H_2O}\sim 10^{-10}$ at high temperatures
($T \sim 3000$~K), produced by an ion-molecule reaction chain. A recent paper by Meijerink et
al. (2011) shows that water production can be enhanced by an order of
magnitude in these warm atomic environments, when the grains are
stripped from their ice layers, as is possibly the case in violent
environments around an AGN.

As we have seen in the previous sections, the temperatures in the
molecular regions of the mechanically heated PDR can be increased to
temperatures $T >> 100$~K, which is favorable for the production of
water through the neutral reaction chain. 

\begin{figure}[!tbh]
  \centering
  \includegraphics[scale=1.0]{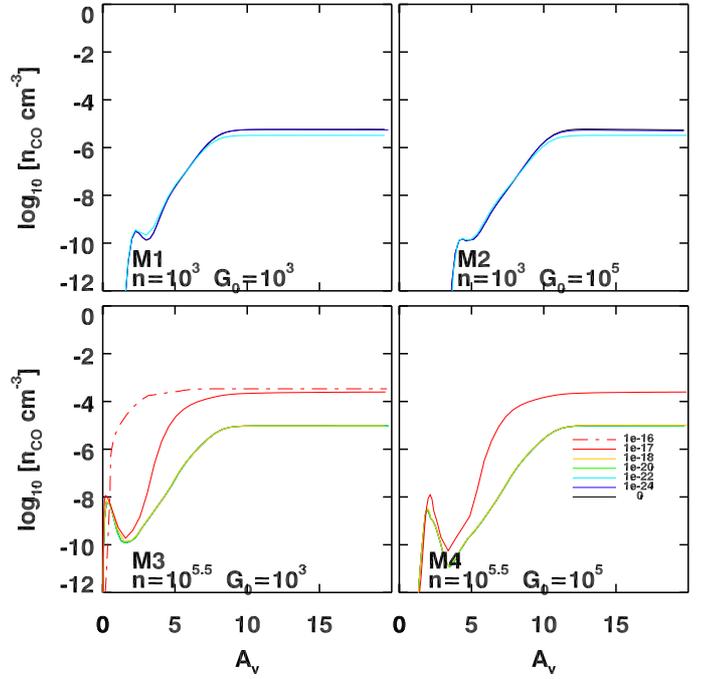}
  \caption{H$_2$O abundance as a function of $A_V$ for selected models.
    \label{fig:selectedModelH2OAbun}}
\end{figure}

\begin{figure}[!tbh]
  \centering
  \includegraphics[scale=2.0]{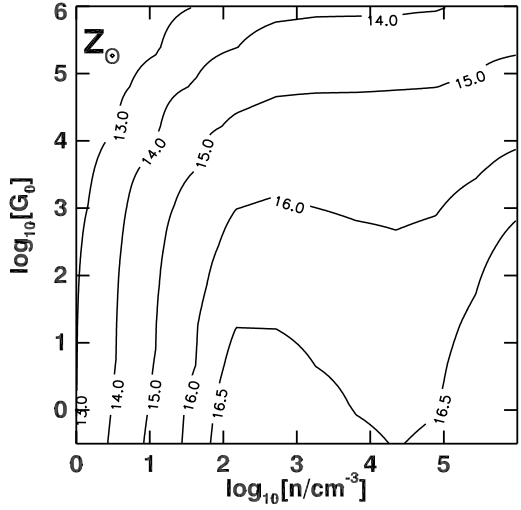}
  \caption{H$_2$O total column density up to $N({\rm H}) = 1.8 \times 10^{22}$~\cms~or $A_V \sim 10$~mag for $Z = Z_\odot$ without mechanical heating. \label{fig:H2OcolDensNoMech}}
\end{figure}

  In Fig.\,\ref{fig:selectedModelH2OAbun} the abundance of H$_2$O
  is shown as a function of depth. The low-density models are
  marginally affected as long as $\Gamma_{\rm mech} <
  10^{-20}$~\ecs~. For higher mechanical heating rates the H/H$_2$
  transition does not occur, thus H$_2$O does not form. However, the
  impact on the abundance of H$_2$O is more significant for the the
  high-density (M3 and M4), where the abundance of H$_2$O is enhanced
  by at least one order of magnitude when $\Gamma_{\rm mech} >
  10^{-17}$~\ecs~. The reference grid for the column density of
  H$_2$O at solar metallicity (without mechanical) heating is shown in
  Fig.\,\ref{fig:H2OcolDensNoMech}. In comparing this reference grid
  to Figs.\,\ref{fig:H2OcolDensMech1}, \ref{fig:H2OcolDensMech2} and
  \ref{fig:H2OcolDensMech0.5} (corresponding to grids for lower and
  higher metallicities), the column densities range between $N({\rm H_2O})\sim 10^{14}$ (very low
densities, $n < 10^1$~\cmt~) and up to $10^{16.5}$\,\cms~($n > 10^2$~\cmt~). These column 
densities increase when the gas density if close to $n = n_c$ especially for $\Gamma_{\rm mech} > 10^{-19}$~\ecs~ where
an optimum of ($N(\rm H_2O)\sim 10^{18}$\,\cms) is reached (see last row of Fig.\,\ref{fig:H2OcolDensMech1}).
When the mechanical heating rate becomes too high (in the regions where $n << n_c$), the column
densities drop again, because there is insufficient H$_2$, which is destroyed through 
collisional dissociation, to form water. Therefore, one finds that this optimum moves from lower to 
higher densities for increasing mechanical heating rates. 

Thus, the high column densities of H$_2$O accompanied by high temperatures 
make this molecule an excellent diagnostic tool for galaxy environments 
where mechanical heating dominates.

\subsubsection{HCN and HNC}

The critical density for the HCN and HNC $J$ = 1-0 collisional
excitation by H is $\sim 1 \times 10^6$ and $\sim 0.5 \times 10^6$
\,\cmt~respectively.  Together with their relatively high abundance,
makes them valuable high gas density tracers. Here we will be first looking 
at the abundances of HNC and HCN, $x_{\rm HNC}$ and $x_{\rm HCN}$ respectively up 
to $A_V = 20$~mag.  Then we will study the effect of 
mechanical heating on the column density ratios of these two species.

\begin{figure}[!tbh]
  \centering
  \includegraphics[scale=1.0]{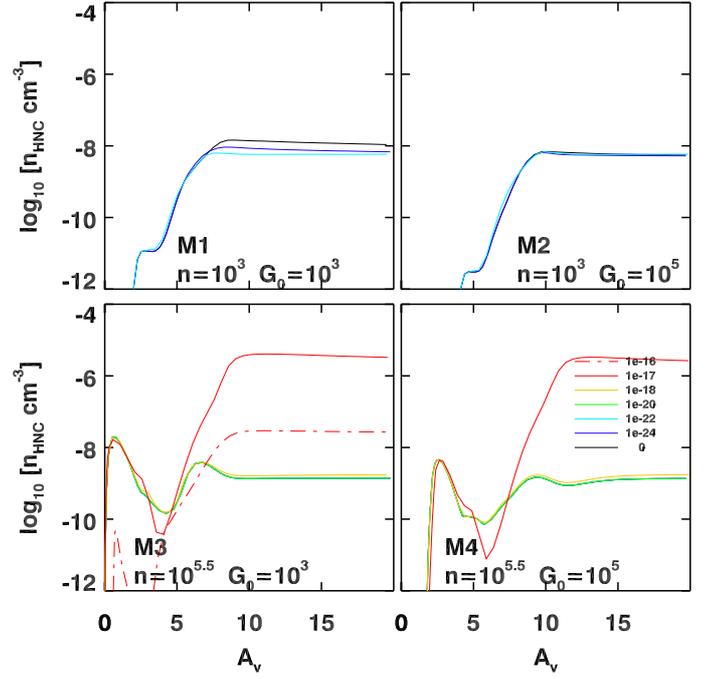}
  \caption{HNC abundance as a function of $A_V$ for selected
    models. In the top panels the curves for $\Gamma_{\rm mech} >
    10^{-22}$\,\ecs~are not shown since the C$^+$/C/CO transition does
    not take place. \label{fig:selectedModelHNCAbun}}
\end{figure}

\begin{figure}[!tbh]
  \centering
  \includegraphics[scale=1.0]{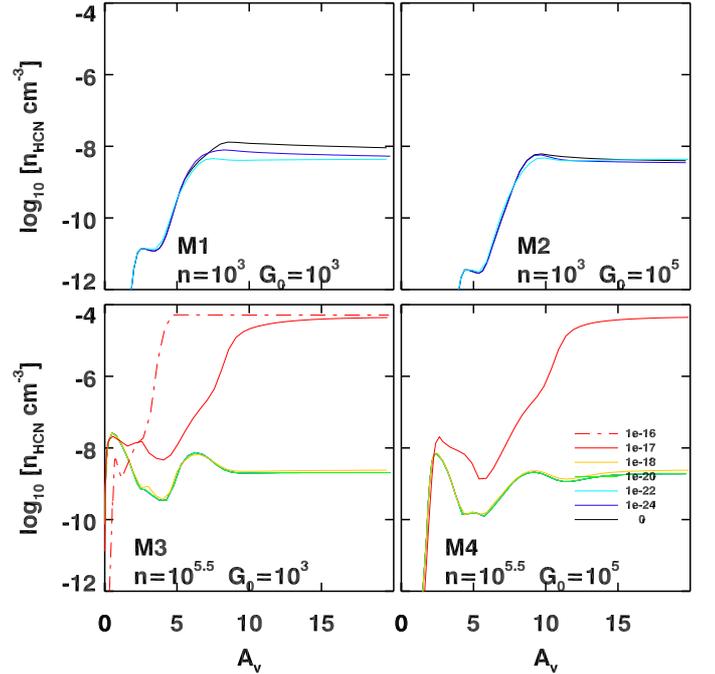}
  \caption{HCN abundance as a function of $A_V$ for selected
    models. The same convention of
    Fig.\,\ref{fig:selectedModelHNCAbun} are
    used. \label{fig:selectedModelHCNAbun}}
\end{figure}

Although included for completeness in
Figs.\,\ref{fig:selectedModelHNCAbun} and
\ref{fig:selectedModelHCNAbun}, the low-density models M1 and M2 are
of no physical interest, since they refer to densities at least two
order of magnitude below the critical density of HCN and HNC. The more
relevant high-density models M3 and M4 show (in contrast to
the low-density models), an increase by four orders of magnitude 
in $x_{\rm HCN}$ when $\Gamma_{\rm mech}$ exceeds $10^{-18}$\,\ecs; whereas an
increase of three orders in $x_{\rm HNC}$ is observed. The increase in the 
abundance of HCN was also noted by \cite{loenen2008} and \cite{meijerink11}. Mechanical
heating causes the equilibrium temperature, in the molecular region, to increase from 100 K, 
when $\Gamma_{\rm mech} = 10^{-18}$\,\ecs, to $\sim 900$ K when 
$\Gamma_{\rm mech} = 10^{-16}$\,\ecs~(see Fig.\,\ref{fig:selectedModelGasTemp}).
Consequently HNC is converted more effectively to HCN via the reaction HNC + H
$\rightarrow$ HCN + H whose activation barrier of $\sim$ 200 K is surpassed \citep{schilke92}. 
This explains the drop in the fractional abundance of HNC with respect to HCN with $x_{\rm HNC} / x_{\rm HCN} \sim 0.1$.

The abundance of HCN and HNC increases by few orders of magnitude with increasing 
$\Gamma_{\rm mech}$~. This increase is because of the enhanced production of 
both species via the neutral-neutral reaction H$_2$ + CN $\rightarrow$ HCN + H. 
Also HCN and HNC are formed with equal probability through 
recombination reactions of HCNH$^+$, which has been also noted by \cite{aalto2005}. 
These two processes are more dominant than the reaction HNC + H $\rightarrow$ HCN + H, which is
just a re-shuffling the bonds of HNC that leaves $x_{\rm HCN} + x_{\rm HNC}$ constant. 
However HNC + H $\rightarrow$ HCN + H is responsible for the bias in the abundance of HCN 
with respect to HNC. 

\begin{figure}[!tbh]
  \centering
  \includegraphics[scale=2.0]{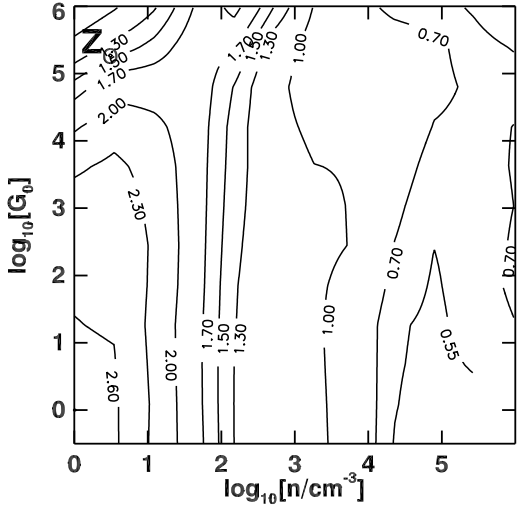}
  \caption{HNC/HCN integrated column density ratios (integrated up to
    $N({\rm H}) = 1.8 \times 10^{22}$\,\cms~or $A_V \sim 10$\,mag for $Z =
    Z_\odot$) without mechanical heating.  The column density ratio
    decreases from left to right. For dense molecular clouds, the
    ratio is $\sim$0.7 and it is almost constant as a function of
    $G_0$. \label{fig:HNCHCNnoMechlDens}}
\end{figure}

Now we look at the column density ratio $N$(HNC)/$N$(HCN) in order to gauge the
effects of mechanical heating in the $G_0 - n$ parameter space. These 
are computed by integrating $x_{\rm HNC}$\, and $x_{\rm HCN}$\, up to 
$N({\rm H})=1.8 \times 10^{22}$\,\cms~and taking their ratio. The same column 
density is considered for the other metallicities when the ratio $N$(HNC)/$N$(HCN)
is computed. In Fig.\,\ref{fig:HNCHCNnoMechlDens} 
the column density ratio of HNC to HCN is shown for the whole grid with no 
mechanical heating (pure PDR).  Since HCN is a good tracer
for dense molecular clouds with $n \ge 10^4$\,\cmt~\citep{solomon1992-1}, we
focus on that part of the parameter space for this diagnostic. In the
absence of mechanical heating, in this range of density, $N$(HNC)/$N$(HCN) $<$ 1
for all $G_0$. When mechanical heating is added, specifically for
$\Gamma_{\rm mech}$ larger than $10^{-19}$\,\ecs, the column density ratio
starts decreasing even more with almost no dependence on $G_0$. For instance,
in the upper right corner $N$(HNC)/$N$(HCN) decreases from 0.7 to 0.55 
(see Fig.\,\ref{fig:HNCHCN-colDensRatio1}). This decrease affects
regions where $\log_{10}n = \log_{10}n_c + 1.0$. At $\log_{10}n = \log_{10}n_c$ 
(dashed vertical lines in Fig.\,\ref{fig:HNCHCN-colDensRatio1}) 
$\Gamma_{\rm mech} = \Gamma_{\rm surface}$ as described in Sect.\,\ref{subsec:surfTemp}, 
so mechanical heating dominates everywhere 
in the cloud. Whereas at $\log_{10}n = \log_{10}n_c + 1.0$ mechanical heating is
up to 100 times smaller than the total heating at the surface of the pure PDR
case. For the largest $\Gamma_{\rm mech}$ considered $10^{-16}$\,\ecs~the 
$N$(HNC)/$N$(HCN) decreases below 0.1 over the whole parameter space (see
Fig.\,\ref{fig:HNCHCN-colDensRatio1}). This is also observed for the other 
metallicities in Figs.\,\ref{fig:HNCHCN-colDensRatio0.5} and 
\ref{fig:HNCHCN-colDensRatio2}.

Thus, we find that the ratio $N$(HNC)/$N$(HCN) is sensitive to low mechanical heating rates,
a factor of a hundred smaller than the total heating at the surface of a pure
PDR. For higher mechanical heating rates that are 20 times smaller than
those at the surface, $N$(HNC)/$N$(HCN) decrease at least by a factor of two.

\subsubsection{HCN/HCO$^+$}

%%% 13 
HCO$^+$ is another high density gas tracer with $n_{cr} \sim
10^5$\,\cmt~for $J$=1-0 transition. It has been suggested that in AGNs, where X-ray heating
dominates, HCN is less abundant than HCO$^+$ \citep{lepp1996},
suggesting that the HCN/HCO$^+$ ratio is a diagnostic for the
determination of the physical nature of a source object.  A detailed
study of the effect of cosmic rays \citep{meijerink11} shows that the
abundance of HCO$^+$ is inhibited by high cosmic ray rates.  Since
cosmic rays have a large penetrating depth, they may raise the
temperature of clouds up to large column densities. Since
$\Gamma_{\rm mech}$ is added uniformly throughout the slab and the
temperatures are likewise enhanced, a behavior similar to that caused
by cosmic rays should be seen.  In a similar fashion as for HNC and HCN, 
we look at abundances and column density ratios with and without 
mechanical heating to see how it affects the $N$(HCN)/$N$(HCO$^+$) ratios.

\begin{figure}[!tbh]
  \centering
  \includegraphics[scale=1.0]{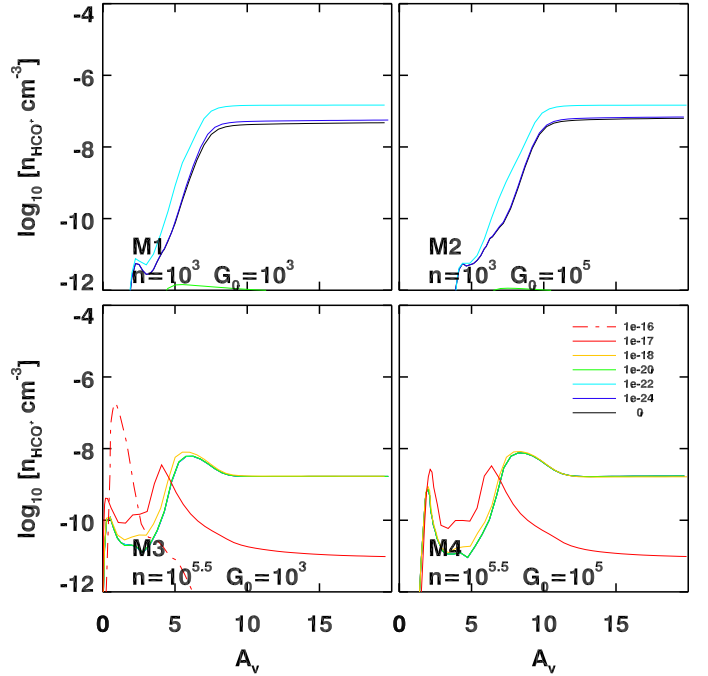}
  \caption{HCO$^+$~abundance as a function of $A_V$ for selected models. \label{fig:selectedModelHCOPAbun}}
\end{figure}

In the low- density models (top row of Fig.\,\ref{fig:selectedModelHCOPAbun}) 
a slight increase in the abundance of HCO$^+$ is observed up to 
$\Gamma_{\rm mech} = 10^{-22}$\,\ecs.  This is caused by a slightly slower 
destruction rate of HCO$^+$ through the reaction HCO$^+$ + e$^-$ 
$\rightarrow$ CO + H. For $\Gamma_{\rm mech} > 10^{-22}$\,\ecs, the two orders 
of magnitude drop in the abundance of HCO$^+$ in the molecular region is 
caused mainly by ion-neutral reactions with H$_2$O, HCN (which we 
demonstrated to increase in the previous section) and C.  The abundance
of HCO$^+$ is unaffected by $\Gamma_{\rm mech}$ in M3 and M4 unless it is larger than
10$^{-17}$\,\ecs.  For that amount of mechanical heating $x_{{\rm HCO}^+}$
becomes $\sim 10^{-11}$ beyond the radical region. In this case, the 
  reactions H$_2$O + HCO$^+$ $\rightarrow$ CO + H$_3$O$^+$ and HCN + HCO$^+$
$\rightarrow$ HCNH$^+$ + CO dominate (again because of enhanced abundances 
of H$_2$O and HCN) in destroying HCO$^+$ by two orders
of magnitude compared to the other reactions. However
an increase of one to four orders of magnitude in $x_{{\rm HCO}^+}$ is
observed for $A_V < 5$\,mag. This contributes significantly to the column 
density of HCO$^+$ in clouds with low $N({\rm H})$, which could make a significant
impact on the interpretation of line ratios for clouds with small
column densities as opposed to those with high $N({\rm H})$.

\begin{figure}[!tbh]
  \centering
 \includegraphics[scale=2.0]{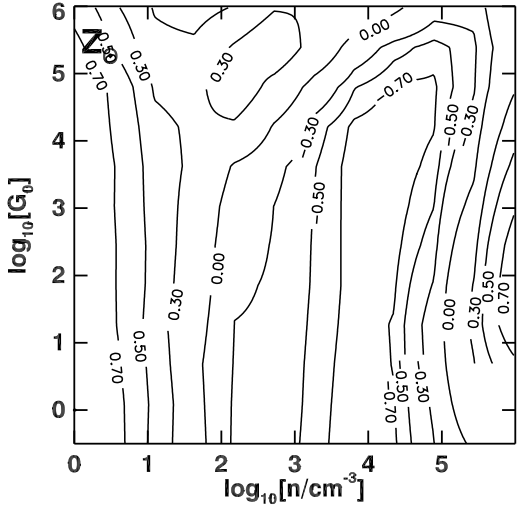}
  \caption{HCN/HCO$^+$ integrated column density ratios (integrated up to $N({\rm H}) = 1.8 \times 10^{22}$~\cms~or $A_V \sim 10$~mag for $Z = Z_\odot$) without mechanical heating. \label{fig:HCNHCOPnoMechlDens}}
\end{figure}

Looking at the column density ratios of HCN to HCO$^+$, $N$(HCN)/$N$(HCO$^+$), we see in Fig.\,\ref{fig:HCNHCOPnoMechlDens} a non-trivial dependence. 
The ratio varies from $10^{-0.7}$ (in the high-density and
high-UV flux zone) to $10^{0.7}$ (in the high-density, moderate- to low-UV
flux zone).

An interesting feature of this column density ratio, quite different from the 
HNC/HCN situation, is the fact that the $N$(HCN)/$N$(HCO$^+$) ratio increases to $\sim 10$ and 
$\sim 10^3$, as opposed to a decrease of {\it only} a factor of two in the ratio 
$N$(HNC)/$N$(HCN) as a function of increasing $\Gamma_{\rm mech}$~.  For high densities when 
$\Gamma_{\rm mech} > 10^{-20}$\,\ecs~the contours become 
almost parallel for $ \log_{10}n_c < \log_{10}n < \log_{10}n_c + 1.0$. This 
implies that the $N$(HCN)/$N$(HCO$^+$) ratio does not depend on $G_0/n$ (as far as  
this ratio is concerned), but is dominated by mechanical heating down to 
values $\sim 100$ times smaller than the surface heating of a pure
PDR. This is shown in Fig.\,\ref{fig:HCNHCOPcolDensRatio1} where 
$N$(HCN)/$N$(HCO$^+$) ranges between $10$ and $10^4$ in the density range 
mentioned. Similar behavior is seen for the other metallicities considered.

This, and the fact that $N$(HCN)/$N$(HCO$^+$) is very
sensitive to very small values of mechanical heating makes this ratio
an excellent diagnostic for mechanically heated regions.

\section{Conclusion and discussion}

The goal of this paper is to assess when we should take into account
mechanical heating in modeling the observed line emission properties
of the interstellar medium in PDRs. We have calculated the equilibrium
thermal and chemical properties of interstellar clouds as 1D
semi-infinite slabs for a large range of hydrogen gas densities ($1 < n <
10^6$~cm$^-3$) and radiation fields ($0.3 < G_0 < 10^6$). As a result,
we covered more or less {\it all} conditions physically relevant for
the ISM, from the diffuse ISM in the outskirts of quiescent galaxies
($G_0 < 1$, $n = 1-100$\,\cmt), to the dense, UV-exposed environments
($G_0 \sim 10^5, n \sim 10^{5.5}$\,\cmt) of active galaxy centers. We
have also explored three levels of heavy-element abundance, $Z=0.5$,
1.0 and 2.0$Z_{\odot}$, thus considering the low metallicities of
dwarf galaxies as well as the enhanced metallicities in galaxy
centers. We have done this by adding a mechanical heating term
resulting from dissipating turbulence of supernovae. The heating rates
 included are between $10^{-24} < \Gamma_{\rm mech} < 10^{-16}$\,\ecs, which
 corresponds to star-formation rates ranging from ${\rm SFR} \sim 0$ to 
$\sim 1000$~M$_{\odot}$~yr$^{-1}$ under the assumptions explained in 
section 2.1.  We first explored the effect on the thermal balance at the 
surface and deep into the cloud at high column densities, and then the 
chemistry of a few commonly observed species in the ISM of galaxies, namely, 
CO, H$_2$O, HCN, HNC, and HCO$^+$. We demonstrated that mechanical heating 
radically changes the resulting chemical and thermal structure. The following
conclusions can be drawn :
 \newline
 (i) The equilibrium temperature in a PDR is significantly altered,
 i.e., increased by an order of magnitude, for ratios $\Gamma_{\rm
   mech}/n^{1.5} \sim 10^{-25}$ erg cm$^{-4.5}$ s$^{-1}$. This is a lower bound in the sense
 that when $\Gamma_{\rm mech}/n^{1.5} = 10^{-25}$ the temperature
 inside the cloud (even for the densest clouds $n \sim 10^6$~\cmt)
 increases by a factor of 10.  Thus, $\Gamma_{\rm mech}$ {\it must}
 be taken into account while solving for the equilibrium state of a
 PDR.  In translating this relation into one relating gas density to
 the star formation rate, it can be expressed as ${\rm SFR}/ n^{3/2} =
 7 \times 10^{-6}$.  For higher values the temperature increases even
 more. Also, the H/H$_2$ transition may not occur for very high
 mechanical heating rates ($\Gamma_{\rm mech}/n^{1.5} \gg 10^{-25}$),
 especially for the low density clouds with $n \sim 10^3$\,\cmt.
 \newline
 (ii) Mechanical heating rates that are two orders of magnitudes
 smaller than the surface heating nevertheless are high enough to
 increase the temperature by an order of magnitude in low-density
 clouds and up to a factor of two in high-density clouds $n \sim
 10^6$\,\cmt. For instance, inclusion of a mechanical heating rate of
 $10^{-18}$~\ecs~ in the high-density cloud models M3 and M4 of
 Meijerink \& Spaans (2005) increases the temperatures inside the
 clouds from $T \sim 30$~K to $\sim 100$~K and from $T \sim 70$~K to
 $\sim 110$~K, respectively. Such heating rates $\Gamma_{\rm mech}$
 are expected to occur in star-burst galaxies, and correspond to an SN
 rate of approximately 0.3 per year and an SFR rate of 9.4
 M$_\odot$~yr$^{-1}$ (\cite{loenen2008}).
 \newline
 (iii) We observe that $N({\rm CO})$ is easily doubled when mechanical
 heating is included in modeling PDRs, especially for gas with $n
 \sim 10^3$\,\cmt.  As such, mechanical heating should be considered,
 when interpreting observed CO data.  We also observe that the implied
 column density strongly depends on metallicity. This column density 
 decreases from $~10^{18}$~\cms~ at solar metallicity to $~10^{17}$~\cms~
 at half-solar metallicity; whereas it increases up to $~10^{19}$~\cms~
 at double-solar metallicity.  This is particularly interesting for galaxy
 centers where the metallicity is expected to be higher than that of the solar
 neighborhood.
 \newline
 (iv) Our results show that the brightness of emission lines observed
 in galaxy centers may at least partly be due to high mechanical
 heating rates. The models with $Z = 2Z_{\odot}$ require higher column
 densities, and temperatures are enhanced by a factor 2 to 10.  It
 also suggests that CO column densities in low-metallicity dwarf galaxies
 are even lower than assumed otherwise. Mechanical heating excites the
 gas to high temperatures, but the column density $N({\rm CO})$ is
 significantly lower (by almost an order of magnitude) compared to $Z =
 Z_{\odot}$). In addition to that, the relative gas fraction is lower
 in dwarf galaxies, as supernovae easily disperse gas from the
 galaxy.
 \newline
 (v) Like CO, H$_2$O also exhibits an increase in column
 density. The temperature increase induced by mechanical heating
 results in implied column densities $N(\rm{H_2 O})$ three orders of
 magnitude higher (up to $\sim 10^{18}$~cm$^{-2}$) compared to models
 without mechanical heating. This gives H$_2$O approximately the same
 column density as CO, for the highest mechanical heating rate. Based
 on this, H$_2$O lines could also provide an excellent diagnostic in
 combination with CO, when studying mechanical heating in galaxies.
 \newline
 (vi) HCN, HNC, HCO+ line ratios have been used to discriminate
 between PDRs and XDRs \citep{meijerink2007-1, loenen2008}. Although
 we do not compute line intensity ratios in this paper, we have
 studied the signature of mechanical heating on the integrated column
 density ratios of HNC/HCN and HCN/HCO+. We find that model HNC/HCN
 column density ratios decrease by a factor of at least 5 for gas
 densities $n > 10^5$~cm$^{-3}$, when $\Gamma_{\rm mech} >
 10^{-17}$~erg~cm$^{-3}$~s$^{-1}$. This effect is mitigated in clouds
 subjected to lower mechanical rates, but only at densities $n <
 10^4$~cm$^{-3}$. HCN and HNC do not produce significant line emission
 at these densities as they are well below the critical density for
 excitation. On the other hand, the HCN/HCO$^+$ column density ratio
 strongly depends on mechanical heating. This ratio increases from
 $\sim 10$ to $\sim 100$, for dense molecular clouds ($n >
 10^5$~cm$^{-3}$), when $\Gamma_{\rm mech} =
 10^{-19}$~erg~cm$^{-3}$~s$^{-1}$ and up to $10^4$ for $\Gamma_{\rm
   mech} = 10^{-18}$~erg~cm$^{-3}$~s$^{-1}$ for $Z = 0.5$, 1.0,
 $2Z_\odot$. In general for these conditions the column densities of
 HCO$^+$ decrease from $\sim 10^{14}$\,\cms~to $\sim 10^{13}$\,\cms,
 those for HCN increase from $\sim 10^{13}$\,\cms~to $\sim
 10^{16.5}$\,\cms, whereas $N({\rm HNC})$ drop to $\sim
 10^{13}$\,\cms~from $\sim 10^{14}$\,\cms. The abundance of HCO$^+$
 drops by at least three orders of magnitude at column densities
 higher $N_{\rm H} > 1.8 \times 10^{21}$~\cms~for the aforementioned
 $\Gamma_{\rm mech}$.  In galaxy centers, higher $A_V$ are easily
 possible compared to $Z = Z_\odot$. This translates into higher
 $N(\rm H)$, which in turn would imply a smaller $N({{\rm HCO}^+})$
 compared to $N({\rm HCN})$. Hence resulting in even higher HCN/HCO+
 column density ratio.  The lack of HCO$^+$ in combination with bright
 HCN line emission might be useful in tracing regions with high
 mechanical heating rates.

The integrated column density ratios do not translate directly into
line intensity ratios. Instead, one should compute these ratios by
solving the radiative transfer equations. We will consider predictions
for the actual line intensities of HCN, HNC and HCO$^+$ in the next
paper, where we will also assess time-dependent effects. The
equilibrium treatment is certainly valid for the low-metallicity dwarf
galaxies, and the benchmark-metallicity of solar neighborhood galaxy
discs, but this may not be the case for the high-metallicity galaxy
centers, where the chemical, thermal, and dynamical timescales are
quite comparable in the inner $< 100$~pc, especially at densities $n <
10^3$~\cmt. For example, the formation timescale of H$_2$
\citep{hollenbach99} is $< 5$ Myr at an ambient gas density $n \sim
10^3$~\cmt~and a kinetic temperature of $T \sim 100$~K.  The stellar
orbital periods (which stir up the ISM) are comparable to or shorter
than this, namely between 0.1 and 15 Myr. Hence, a time-dependent
treatment is necessary, and we aim to accomplish this in upcoming
papers by adapting the current PDR-XDR code. We will allow for
time-dependent chemical evolution, and couple it with an SPH code
through AMUSE. As such, the chemistry, gas and stellar dynamics are
being evolved self-consistently.

\begin{acknowledgements}
  M.V.K would like to thank A.G.G.M. Tielens, M. Hogerheijde and Ewine F. van Dishoeck for useful
  discussions about the chemical reaction network. M.V.K is also thankful to V. Icke for thoughtful
  advice about implementing the adaptive slab discretization and E. Loenen for discussions about 
  relating the mechanical heating to the star-formation rate.
\end{acknowledgements}

%\bibliographystyle{aa}
%\bibliography{references}

%----------------------------------------------------------------------------------------------------
%----------------------------------------------------------------------------------------------------
%----------------------------------------------------------------------------------------------------

%----------------------------------------------------------------------------------------------------
%----------------------------------------------------------------------------------------------------
%----------------------------------------------------------------------------------------------------

\begin{appendix}
\end{appendix}

\begin{appendix}

\clearpage
\begin{figure*}[!tbh]
  \centering
  \includegraphics[scale=1.5]{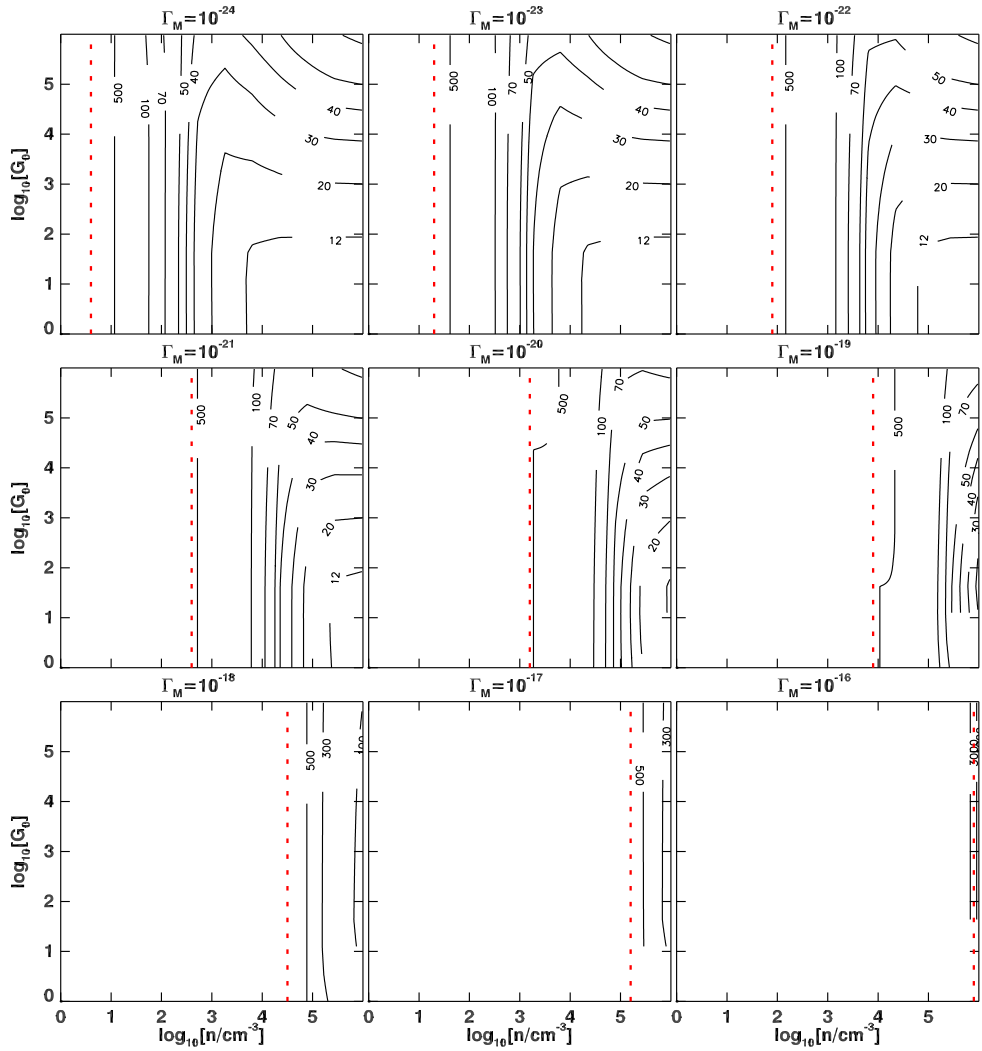}
  \caption{Temperature contours at $A_V = 20$ for different values of mechanical heating. As 
    mechanical heating is increased from $10^{-24}$ to $10^{-16}$, the temperatures increase
    from left to right in the grid as function of increasing mechanical heating. The temperatures
    to the left of the highest temperature contour line are higher than the temperature
    of the contour curve itself, which are usually larger than 10000K. The models in these 
    regions are not very accurate since cooling effects for temperature larger than 10000K are 
    not included properly, so we do not plot the corresponding contours. \label{fig:insideTempWithMech} }
\end{figure*}

\clearpage
\begin{figure*}[!tbh]
  \centering
  \includegraphics[scale=1.5]{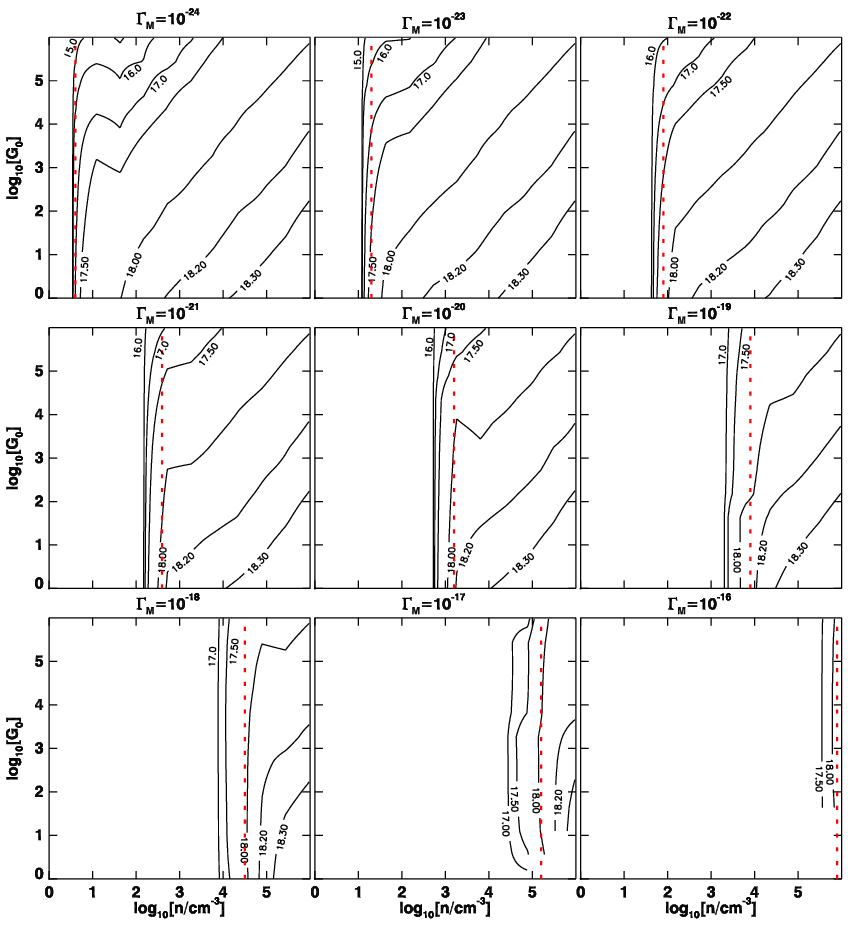}
  \caption{$\log_{10}$ CO total column density (integrated up to $N({\rm H}) = 1.8 \times 10^{22}$\cms~or $A_V \sim 10$mag for $Z = Z_\odot$) for different values of mechanical heating.
    \label{fig:COcolDensMech1}}
\end{figure*}

\clearpage
\begin{figure*}[!tbh]
  \centering
  \includegraphics[scale=1.5]{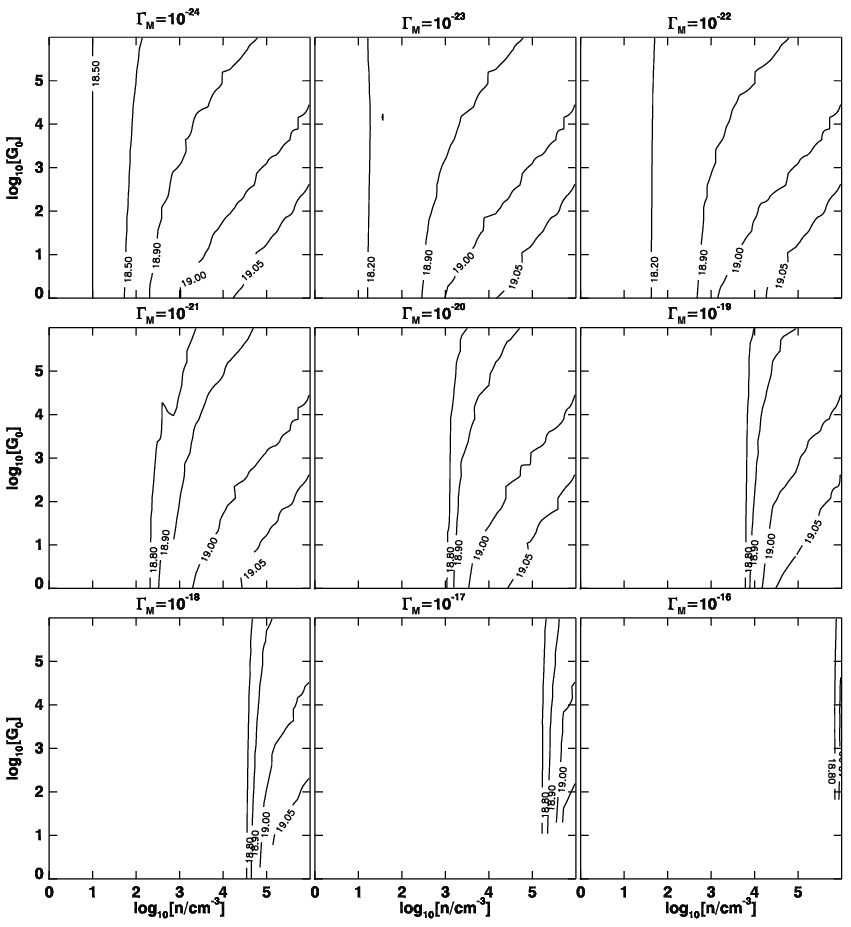}
  \caption{$\log_{10}$ CO total column density (integrated up to $N({\rm H}) = 1.8 \times 10^{22}$\cms~or $A_V \sim 20$mag for $Z = 2 Z_\odot$) as a function of mechanical heating for the double solar metallicity case.\label{fig:COcolDensMech2}}
\end{figure*}

\clearpage
\begin{figure*}[!tbh]
  \centering
  \includegraphics[scale=1.5]{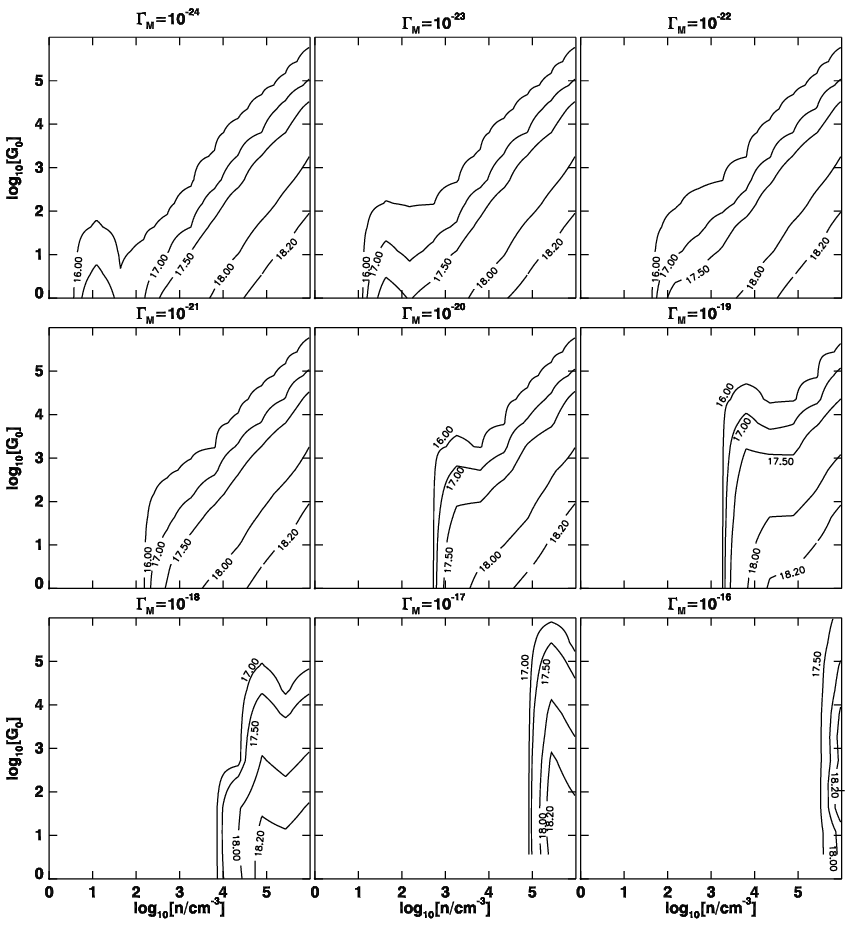}
  \caption{$\log_{10}$ CO total column density (integrated up to $N({\rm H}) = 1.8 \times 10^{22}$\cms~or $A_V \sim 5$mag for $Z = 0.5 Z_\odot$) as a function of mechanical heating for the low metallicity case. \label{fig:COcolDensMech0.5}}
\end{figure*}

\clearpage
\begin{figure*}[!tbh]
  \centering
  \includegraphics[scale=1.5]{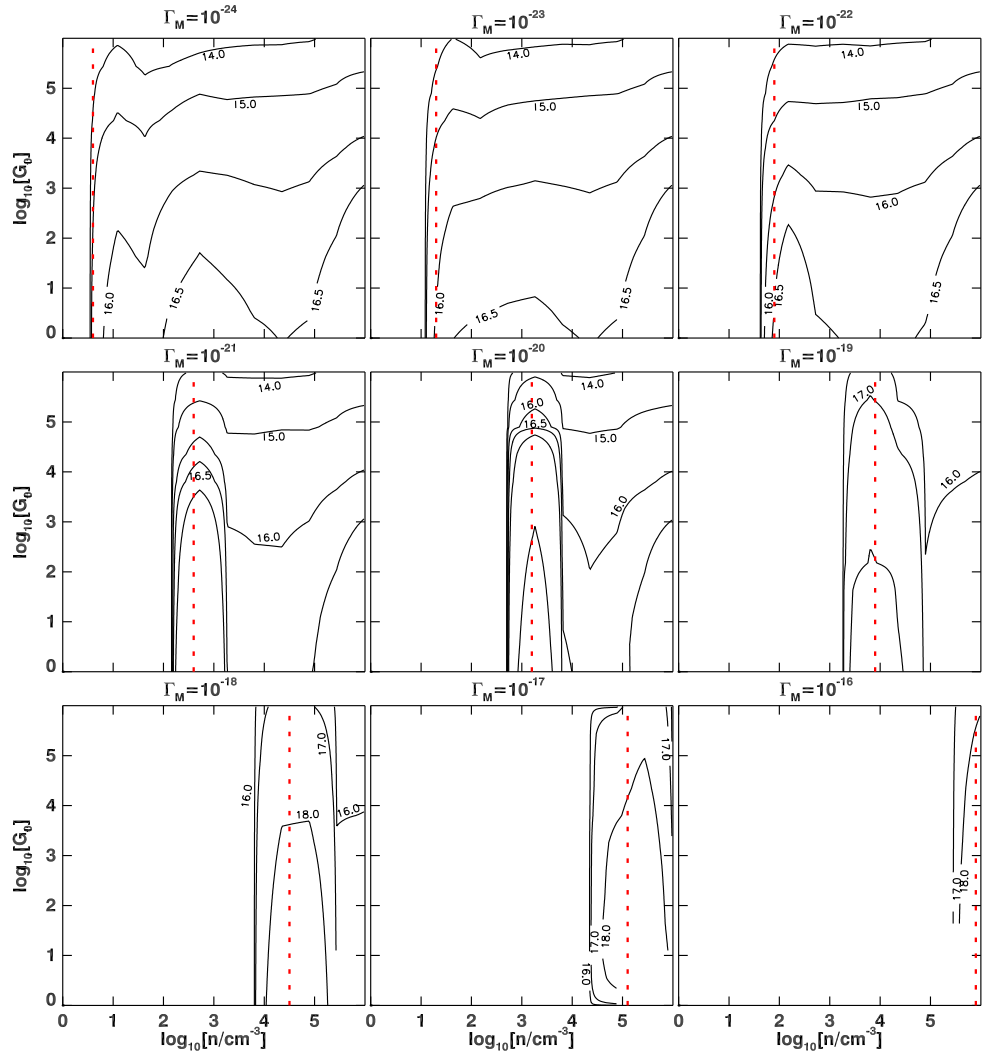}
  \caption{$\log_{10}$ H$_2$O total column density (integrated up to $N({\rm H}) = 1.8 \times 10^{22}$\cms~or $A_V \sim 10$mag for $Z = Z_\odot$) for different values of mechanical heating.
    \label{fig:H2OcolDensMech1}}
\end{figure*}

\clearpage
\begin{figure*}[!tbh]
  \centering
  \includegraphics[scale=1.5]{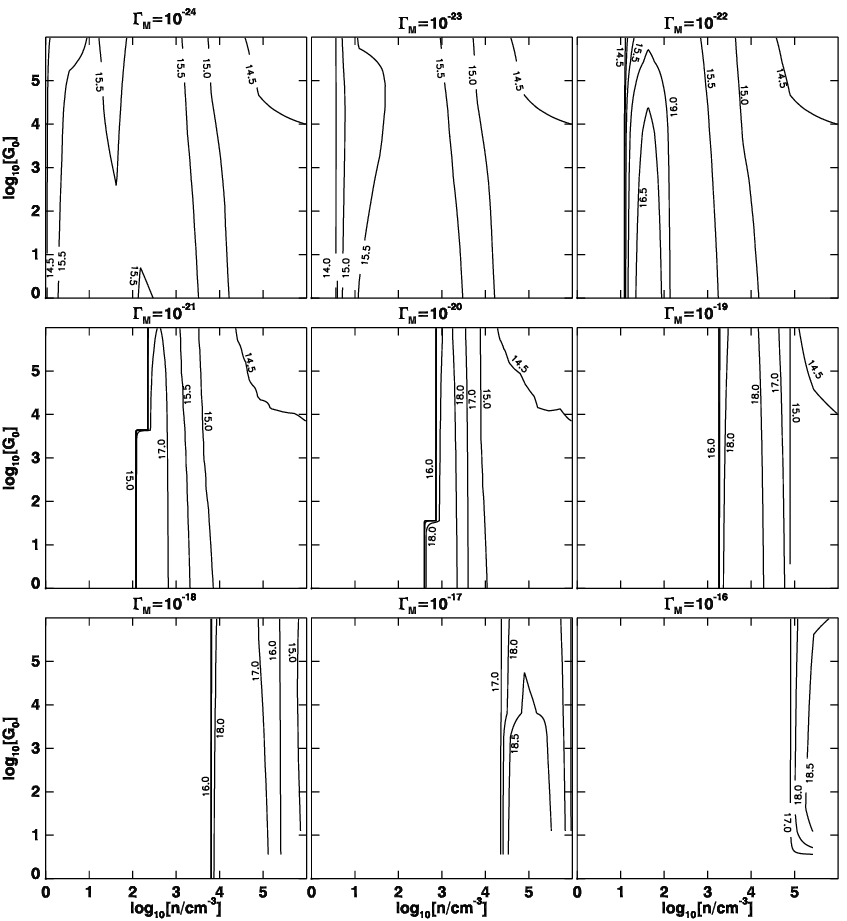}
  \caption{$\log_{10}$ H$_2$O total column density (integrated up to $N({\rm H}) = 1.8 \times 10^{22}$\cms~or $A_V \sim 20$mag for $Z = 2 Z_\odot$) as a function of mechanical heating for the double solar metallicity case.\label{fig:H2OcolDensMech2}}
\end{figure*}

\clearpage
\begin{figure*}[!tbh]
  \centering
  \includegraphics[scale=1.5]{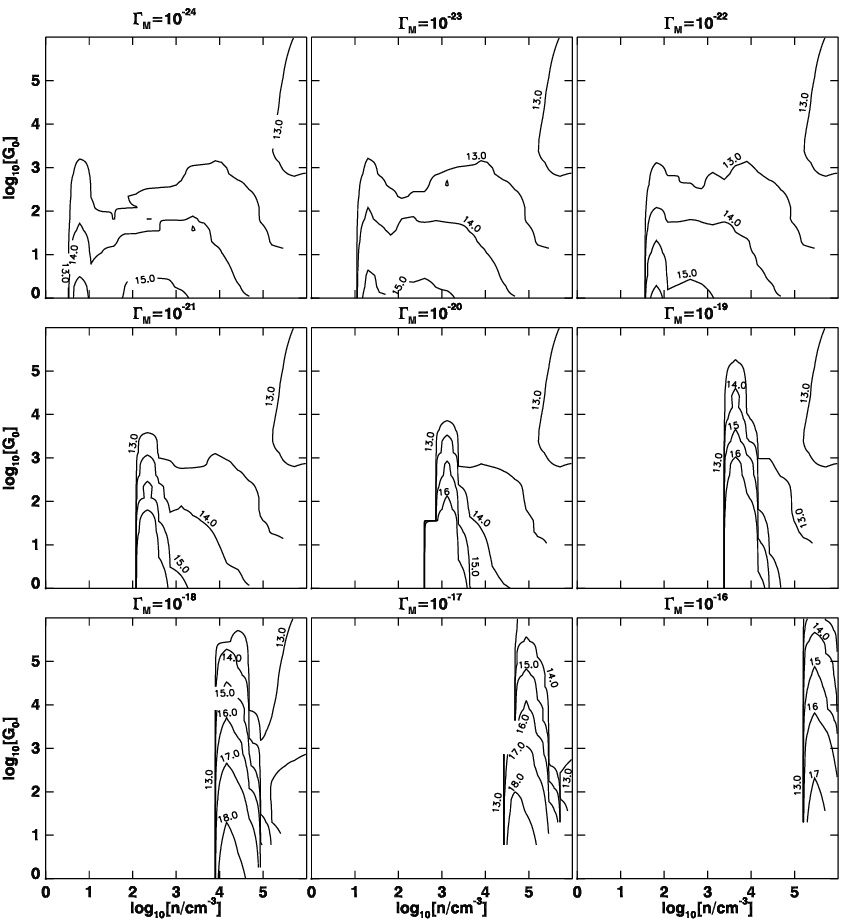}
  \caption{$\log_{10}$ H$_2$O total column density (integrated up to $N({\rm H}) = 1.8 \times 10^{22}$\cms~or $A_V \sim 5$mag for $Z = 0.5 Z_\odot$) as a function of mechanical heating for the low metallicity case. \label{fig:H2OcolDensMech0.5}}
\end{figure*}

\clearpage
\begin{figure*}[!tbh]
  \centering
  \includegraphics[scale=1.5]{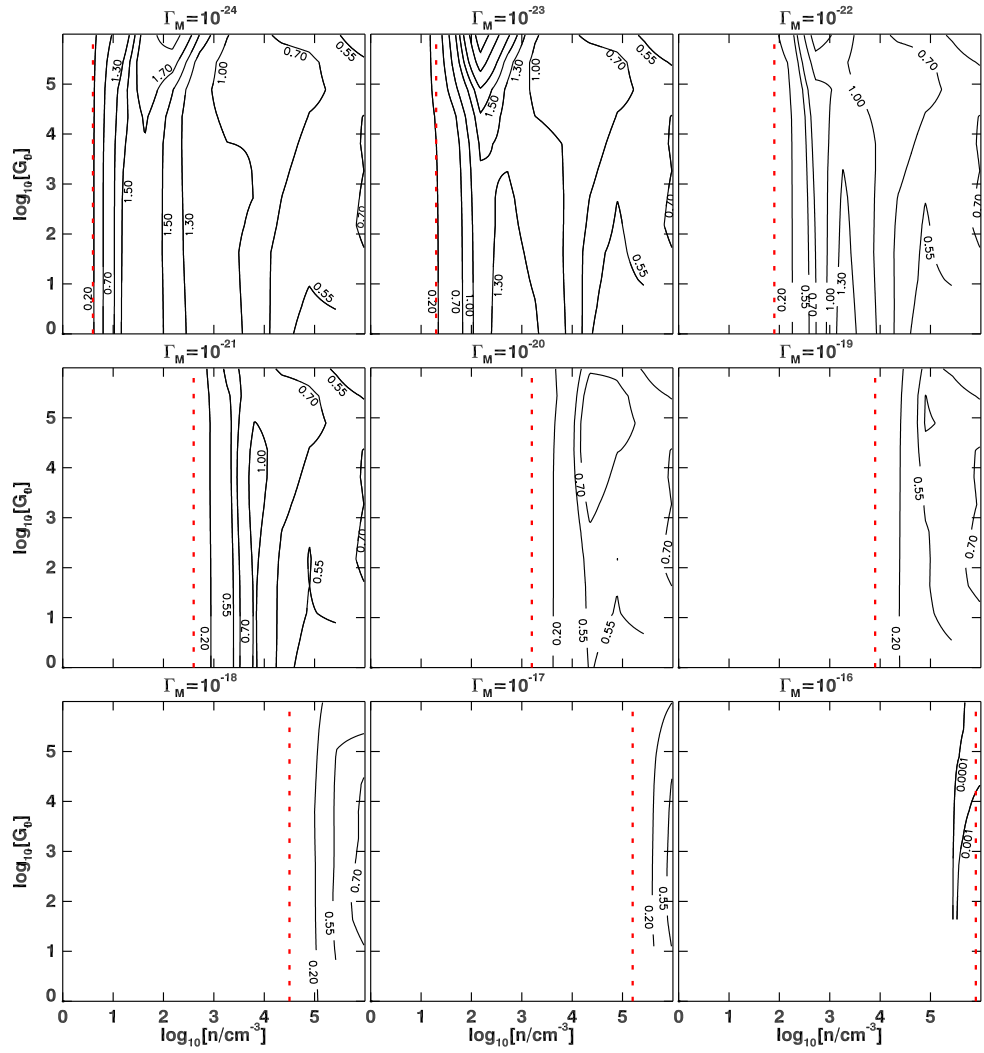}
  \caption{HNC/HCN integrated column density ratios (integrated up to $N({\rm H}) = 1.8 \times 10^{22}$\cms~or $A_V \sim 10$mag for $Z = Z_\odot$) as a function of mechanical heating. \label{fig:HNCHCN-colDensRatio1}}
\end{figure*}

\clearpage
\begin{figure*}[!tbh]
  \centering
  \includegraphics[scale=1.5]{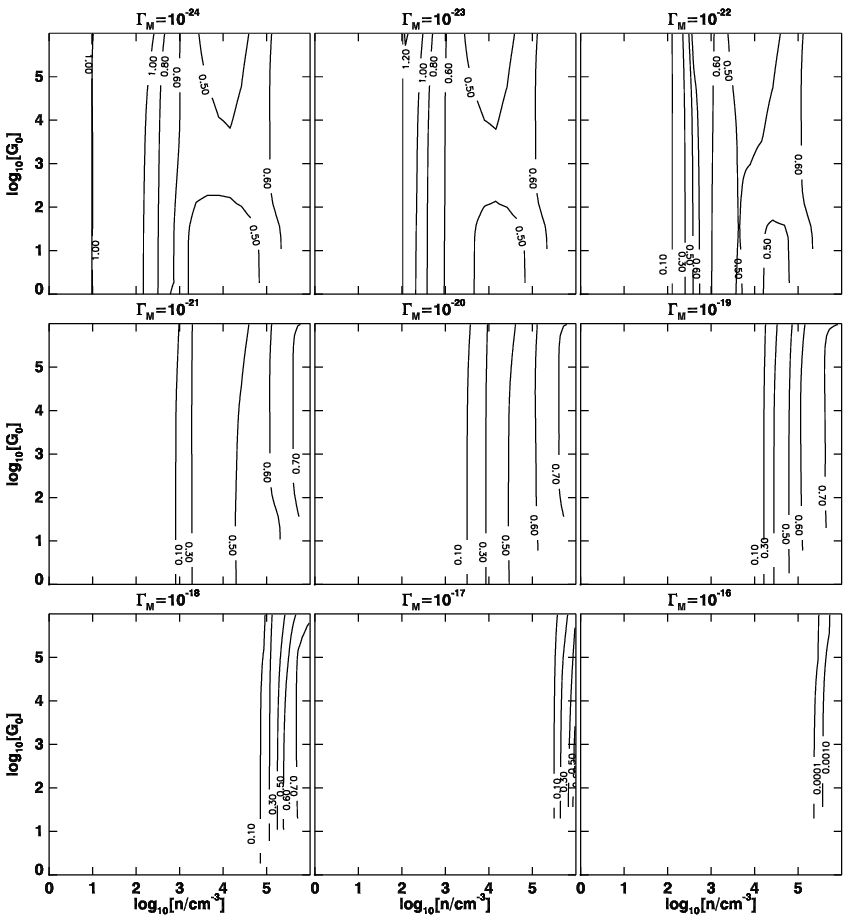}
  \caption{HNC/HCN column density ratios (integrated up to $N({\rm H}) = 1.8 \times 10^{22}$\cms~or $A_V \sim 20$mag for $Z = 2 Z_\odot$) as a function of mechanical heating for the double solar metallicity case. \label{fig:HNCHCN-colDensRatio2}}
\end{figure*}

\clearpage
\begin{figure*}[!tbh]
  \centering
  \includegraphics[scale=1.5]{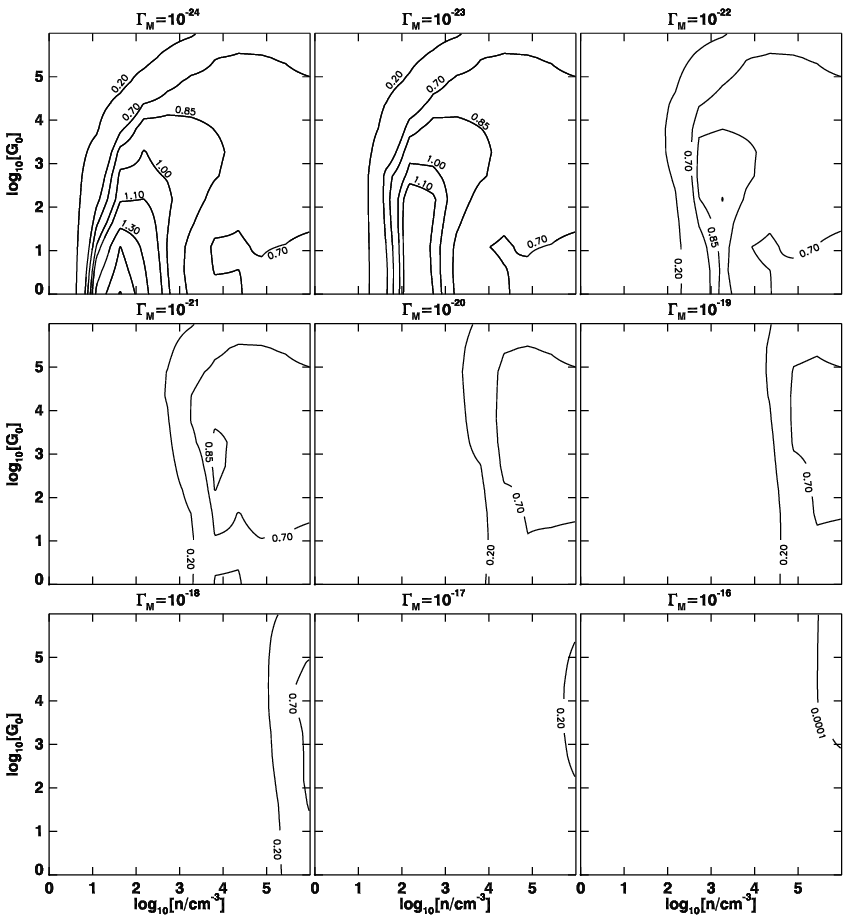}
  \caption{HNC/HCN integrated column density ratios (integrated up to $N({\rm H}) = 1.8 \times 10^{22}$\cms~or $A_V \sim 5$mag for $Z = 0.5 Z_\odot$) as a function of mechanical heating for the low metallicity case. \label{fig:HNCHCN-colDensRatio0.5}}
\end{figure*}

\clearpage
\begin{figure*}[!tbh]
  \centering
  \includegraphics[scale=1.5]{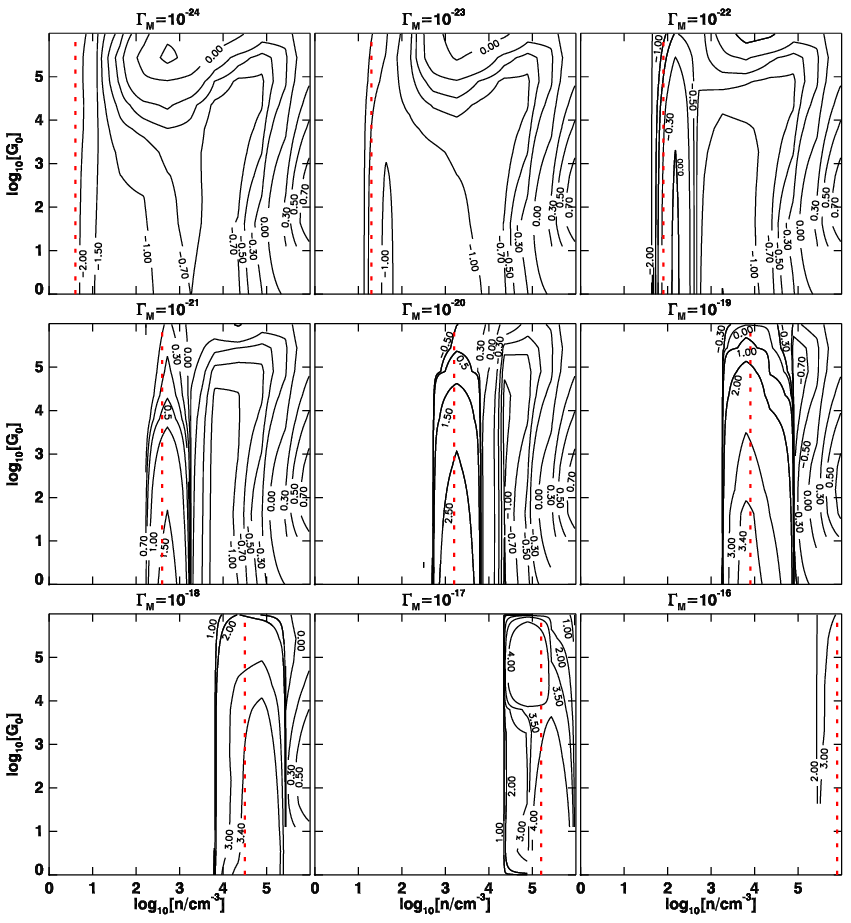}
  \caption{$\log_{10}$  HCN/HCO$^+$ integrated column density ratios (integrated up to $N({\rm H}) = 1.8 \times 10^{22}$\cms~or $A_V \sim 10$mag for $Z = Z_\odot$) as a function of mechanical heating for solar metallicity.\label{fig:HCNHCOPcolDensRatio1}}
\end{figure*}

\clearpage
\begin{figure*}[!tbh]
  \centering
  \includegraphics[scale=1.5]{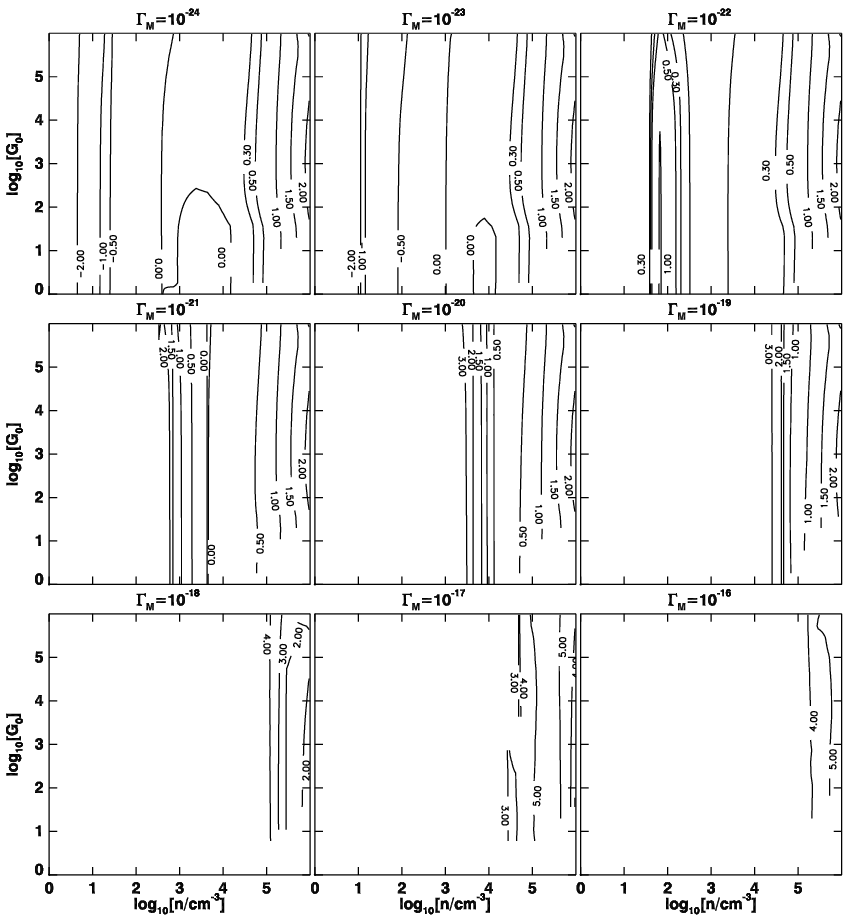}
  \caption{$\log_{10}$ HCN/HCO$^+$ column density ratios (integrated up to $N({\rm H}) = 1.8 \times 10^{22}$\cms~or $A_V \sim 20$mag for $Z = 2 Z_\odot$) as a function of mechanical heating for the double solar metallicity case.\label{fig:HCNHCOPcolDensRatio2}}
\end{figure*}

\clearpage
\begin{figure*}[!tbh]
  \centering
  \includegraphics[scale=1.5]{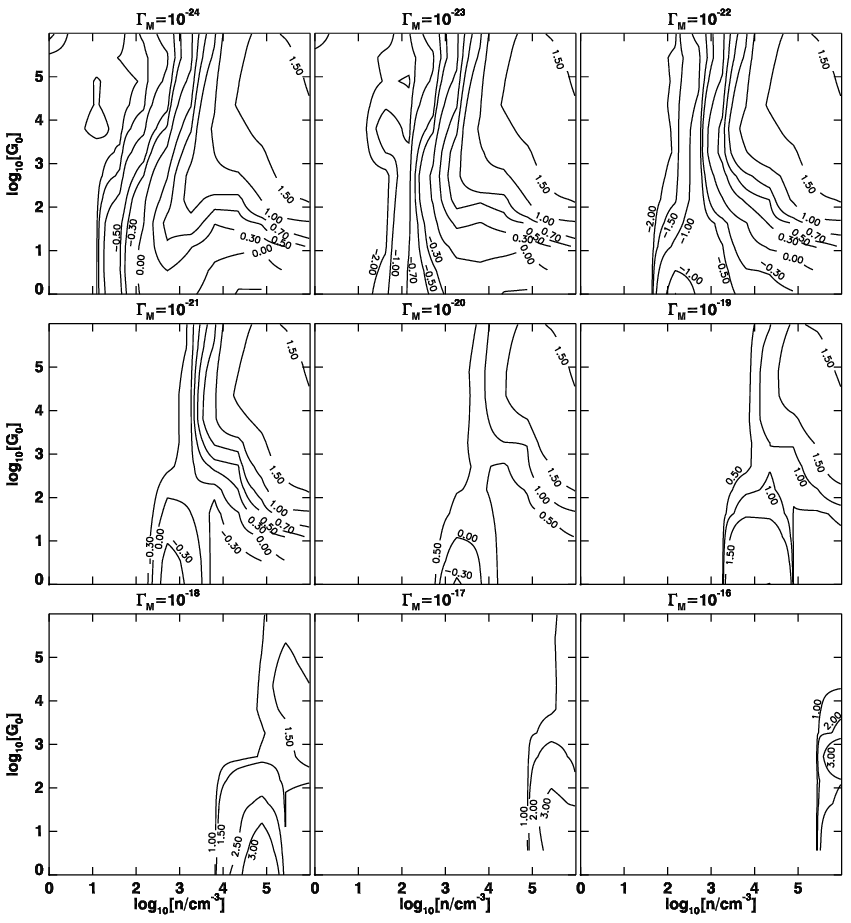}
  \caption{$\log_{10}$ HCN/HCO$^+$ integrated column density ratios (integrated up to $N({\rm H}) = 1.8 \times 10^{22}$\cms~or $A_V \sim 5$mag for $Z = 0.5 Z_\odot$) as a function of mechanical heating for the low metallicity case. \label{fig:HCNHCOPcolDensRatio0.5}}
\end{figure*}

\end{appendix}

\end{document}